\documentclass[a4paper]{article}
\usepackage{latexsym,amssymb,amsmath}
\usepackage{amscd,array,cite,dsfont,enumitem,graphicx,mathtools,multirow,rotating,tensor,times,verbatim,youngtab,placeins}
\RequirePackage{xcolor}
\definecolor{MyDarkBlue}{rgb}{0.15,0.25,0.45}
\usepackage[bookmarksnumbered,bookmarksopen=true,bookmarksopenlevel=1,pdfstartview=FitH, colorlinks=true, linkcolor=MyDarkBlue, citecolor=MyDarkBlue, urlcolor=MyDarkBlue]{hyperref}
\usepackage[all]{hypcap}
\usepackage[margin=3cm]{geometry}
\input xy
\xyoption{all}
\usepackage{tikz-cd}
\newcolumntype{M}[1]{>{$}{#1}<{$}}

\DeclareMathAlphabet\Scr{U}{rsf}{m}{n} \makeatletter
\@addtoreset{equation}{section} \makeatother

\newcommand{\be}{\begin{equation}}
\newcommand{\ee}{\end{equation}}
\newcommand{\bea}{\begin{eqnarray}}
\newcommand{\eea}{\end{eqnarray}}
\newcommand{\ba}{\begin{array}}
\newcommand{\ea}{\end{array}}
\newcommand{\bit}{\begin{itemize}}
\newcommand{\eit}{\end{itemize}}
\newcommand{\ben}{\begin{enumerate}}
\newcommand{\een}{\end{enumerate}}

\allowdisplaybreaks[1]

\DeclareMathOperator{\SO}{SO}

\DeclareMathOperator{\SU}{SU}
\DeclareMathOperator{\Un}{U}

\newcommand{\R}{\mathds{R}}

\newcommand{\Z}{\mathds{Z}}

\newcommand{\sst}[1]{{\scriptscriptstyle #1}}

\def\0{{\sst{(0)}}}
\def\1{{\sst{(1)}}}
\def\2{{\sst{(2)}}}
\def\3{{\sst{(3)}}}
\def\4{{\sst{(4)}}}
\def\5{{\sst{(5)}}}
\def\6{{\sst{(6)}}}
\def\7{{\sst{(7)}}}
\def\n{{\sst{(n)}}}

\begin{document}

\begin{titlepage}
\begin{center}
\hfill Imperial-TP-2021-MJD-01\\

\vskip 1.5cm

{\huge \bf Odd dimensional analogue of the Euler characteristic}

\vskip 1.5cm

{\large L.~Borsten${}^a$,  M.~J.~Duff${}^b{}^c$ and S.~Nagy${}^d$}

\vskip 20pt

{\it ${}^a$Maxwell Institute and Department of Mathematics\\
Heriot-Watt University, Edinburgh EH14 4AS, United Kingdom\\\vskip 5pt

${}^b$Institute for Quantum Science and Engineering and Hagler Institute for Advanced Study, Texas A\&M University, College Station, TX, 77840, USA\\\vskip 5pt
${}^c$Theoretical Physics, Blackett Laboratory, Imperial College London,\\
 London SW7 2AZ, United Kingdom\\\vskip 5pt

${}^d$Queen Mary University of London, 327 Mile End Road, London E1 4NS, United Kingdom}\\\vskip 5pt
\href{mailto:l.borsten@hw.ac.uk}{\ttfamily l.borsten@hw.ac.uk}, ~\href{mailto:m.duff@imperial.ac.uk}{\ttfamily m.duff@imperial.ac.uk}, ~\href{mailto:s.nagy@qmul.ac.uk}{\ttfamily s.nagy@qmul.ac.uk}

\end{center}

\vskip 2.2cm

\begin{center} {\bf ABSTRACT}\\[3ex]\end{center}

When compact 
manifolds  $X$ and $Y$ are both even dimensional, their Euler characteristics obey the K\"unneth formula $\chi(X\times Y)=\chi(X) \chi(Y)$. In terms of the Betti numbers $b_p(X)$, $\chi(X)=\sum_{p}(-1)^p b_p(X)$, implying that $\chi(X)=0$ when $X$ is odd dimensional.
We seek a linear combination of Betti numbers, called $\rho$, that  obeys an analogous formula $\rho(X\times Y)=\chi(X) \rho(Y)$ when $Y$ is odd dimensional. The unique solution is
 $\rho(Y)=-\sum_{p}(-1)^p p b_p(Y)$.  Physical applications include: (1) $\rho \rightarrow (-1)^m \rho $ under a generalized mirror map in $d=2m+1$ dimensions, in analogy with   $\chi \rightarrow (-1)^m \chi $ in $d=2m$; (2) $\rho$ appears naturally in compactifications of M-theory. For example,  the 4-dimensional Weyl anomaly for M-theory on $X^4 \times Y^7$ is given by $\chi(X^4)\rho(Y^7)=\rho(X^4 \times Y^7) $ and hence vanishes when $Y^7$ is self-mirror. Since, in particular, $\rho(Y\times S^1)=\chi(Y)$,  this is consistent with  the corresponding anomaly for Type IIA on $X^4 \times Y^6$, given by $\chi(X^4)\chi(Y^6)=\chi(X^4 \times Y^6)$,
which vanishes when $Y^6$ is self-mirror; (3)  In the partition function of $p$-form gauge fields, $\rho$ appears in odd dimensions as $\chi$ does in even.



\vfill

\end{titlepage}

\newpage \setcounter{page}{1} \numberwithin{equation}{section}

\newpage\tableofcontents

\newpage
\indent

\section{ Introduction} 

The familiar Euler characteristic of a manifold $X$, given by the alternating sum of Betti numbers $b_p(X)$,
\be
\chi(X)=\sum_{p=0}^{d} (-1)^p b_p(X) 
\ee
is identically zero when $d=\dim X=2m+1$. In this paper we argue that in several respects the topological invariant
\be
\rho(X)= -\sum_{p=0}^{d} (-1)^p pb_p(X) 
\ee
is a natural   generalisation of the Euler characteristic that is non-trivial in  odd dimensions:
\begin{enumerate}[label=(\roman*)]
\item{Kunneth formula (\autoref{sec_rho}):}

When closed manifolds  $X$ and $Y$ are both even dimensional, their Euler characteristics obey the non-trivial\footnote{Of course, it is satisfied trivially when at least one manifold is odd dimensional. } K\"unneth formula 
\be
\chi(X\times Y)=\chi(X) \chi(Y)
\label{kunchi}
\ee
whereas $\rho$ obeys an analogous formula when $\dim Y$ is odd
\be
\rho(X\times Y)=\chi(X) \rho(Y).
\label{kunrho}
\ee

Indeed it is the unique (up to trivial scaling and shifts, cf.~\autoref{sec_rho}) linear combination of Betti numbers to do so.
\item{Special case $Y^{(d+1)}=X^d \times S^1$  (\autoref{sec_rho}):}

In this case
\be
\rho(Y)=\chi(X).
\label{special}
\ee
\item{Generalised mirror map  (\autoref{sec_mirror}):}

For $d=2m$, there is a mirror map under which
\be
\chi(X) \rightarrow (-1)^m \chi(X)
\ee
For $d=2m+1$, there is a mirror map under which
\be
\rho(Y) \rightarrow (-1)^m \rho(Y)
\ee
\item{Weyl anomalies in Type IIA and M-theory compactifications  (\autoref{sec_weyl}):}

The topological invariant $\rho$ first made its appearance in the case $d=7$ corresponding to a compactification of M-theory from $D=11$ to $D=4$ spacetime dimensions\footnote{Note, we use $D$ ($d$) to refer to the dimension of a Lorentzian (Riemannian) manifold.}  \cite{Duff:2010ss,Duff:2010vy,Duff:2016whi}, as we now recall in  \autoref{D=11}.

For $d=10$ Type IIA on $X^4(spacetime) \times X^6(internal)$ the $d=4$ on-shell Weyl anomaly ${\cal A}^{\rm W}$  is given by\footnote{Where it is understood that we Wick rotate the spacetime manifold to be Euclidean and assume it is closed. }
\be
\int {\cal A}^{\rm W}=-\frac{1}{24}\chi(X^4)\chi(X^6)=-\frac{1}{24}\chi(X^4 \times X^6)
 \ee
 on using the K\"unneth rule \eqref{kunchi} and hence vanishes when $X^6$ (and therefore $X^{10}$) is self-mirror.
 
 For $d=11$  M-theory on $X^4(spacetime) \times Y^7(internal)$  the $d=4$ on-shell Weyl anomaly ${\cal A}^{\rm W}$ is given by
 \be
\int {\cal A}^{\rm W}=-\frac{1}{24}\chi(X^4)\rho(Y^7)=-\frac{1}{24}\rho(X^4 \times Y^7)
\label{rho7}
\ee
 on using the K\"unneth rule (\ref{kunrho}) and hence vanishes when $Y^7$ (and therefore $X^{11}$) is self-mirror.
 
 By virtue of the special case  (\ref{special}), this is entirely consistent with the equivalence
 \be
 \text{M-theory}~ \text{on}~ X^{10} \times S^1 \equiv \text{IIA}~\text{on} ~X^{10}.
 \ee

\item{Quantum inequivalence of $p$-forms and their duals (\autoref{Z}):}

The partition functions for  a $p$-form gauge field need not be the same as that of its dual $\tilde p=d-p-2$ \cite{Duff:1980qv, Witten:1995gf, Donnelly:2016mlc}. In even dimensions
they differ by Euler characteristic terms. In odd dimensions they are equivalent but $\rho$ still makes an appearance via the logarithmic contribution from the non-zero modes of the Laplacian on $p$-forms 
\be
(-1)^p \rho (X)
\ee
It also appears in the partition function of $BF$ theories and the special cases of odd-dimensional Chern-Simons theories.
\end{enumerate}



\begin{table}[h!]
\[\begin{array}{|l|l|r|r|r|r|}
\hline
\hline
D=11&D=4  &f(\phi)&360A(\phi)&\text{Multiplicity for}~ Y^7&\text{Multiplicity for}~ X^6 \times S^1\\
&&&&&\\
\hline
g_{MN}&g_{\mu\nu}&2&848&b_0&c_0\\
~&{\cal A}_{\mu}&2&-52&b_1&c_0+c_1\\
~&{\cal A}&1&4&-b_1+b_3&-c_0-c_1+c_2+c_3\\
\psi_{M}&\psi_{\mu}&2&-233&b_0+b_1&2c_0+c_1\\
~&\chi&2&7&b_2+b_3&c_1+2c_2+c_3\\
A_{MNP}&A_{\mu\nu\rho}&0&-720&b_0&c_0\\
~&A_{\mu\nu}&1&364&b_1&c_0+c_1\\
~&A_{\mu}&2&-52&b_2&c_1+c_2\\
~&A&1&4&b_3&c_2+c_3\\
&&&&&\\
\hline
\text{total}~f&&&&4(b_0+b_1+b_2+b_3)&4(2c_0+2c_1+2c_2+c_3)\\
&&&&&\\
\text{total} ~A&&&&-(7b_0-5b_1+3b_2-b_3)/24&-(2c_0-2c_1+2c_2-c_3)/24\\
&&&&&\\
~&&&&=-\rho(Y^7)/24&=-\chi(X^6)/24\\
\hline
\hline
\end{array}\]
\caption{Compactification of $D=11$ supergravity to $D=4$ on manifolds $Y^7$ and $X^6\times S^1$ assuming at least one unbroken supersymmetry.  Here, $f(\phi)$ counts the on-shell degrees of freedom and $A(\phi)$ the contribution of a field $\phi$ to the Weyl anomaly coefficient $A$. Note, $b_k:=b_k(Y^7)$ and $c_k:=b_k(X^6)$. }
\label{D=11}
\end{table}

\section{The $\rho$-characteristic}\label{sec_rho}

\subsection{Definition}
Let us  confine our attention to  closed manifolds.
In \cite{Duff:2010ss, Duff:2010vy, Duff:2016whi} it was observed that for a   7-manifold $X$,  the combination of Betti numbers
\be\label{rho0}
\rho(X) = \sum_{p=0}^{3} (-1)^p(7-2p) b_p(X) = 7 b_0 - 5 b_1 +3 b_2 -b_3,
\ee
plays a special role in M-theory compactifications to $D=4$, analogous to that played by Euler characteristic in type II string theory compactifications to $D=4$. 

It obviously generalises to arbitrary odd dimensional manifolds $\dim X = 2m+1$,
\be\label{rho1}
\rho(X) = \sum_{p=0}^{m} (-1)^p(\dim X-2p) b_p(X) ,
\ee
and obeys the K\"unneth-type formula,
\be\label{prod1}
\rho(X \times Y) = \rho(X)\chi(Y),
\ee
where $X$ and $Y$ are even and odd dimensional, respectively.

We then  define $\rho$ in all dimensions, even and odd, by
\be\label{rho2}
\rho(X) := - \sum_{p=0}^{d}(-1)^p p b_p(X).
\ee
 By Poincare duality,
\be\label{bs}
b_p(X)= b_{d-p}(X),
\ee
this yields \eqref{rho1} for $\dim X =2m+1$ and 
\be
\rho(X) = -m \chi(X),
\ee
for $\dim X=2m$. 

Obviously, \eqref{rho2} is not the unique choice yielding \eqref{rho1} in odd dimensions since the Betti numbers are not independent due to Poincar\'e duality. In particular, we may add any amount of $\chi(X)$ while preserving \eqref{rho1}. The specific choice \eqref{rho2} is natural in many regards and  is  reminiscent of the Ray-Singer torsion \cite{RAY1971145}; it shares many of its properties.  

In particular, using the  K\"unneth formula 
\be\label{kunneth}
b_r(X\times Y) = \sum_{p+q=r}b_p(X)b_q(Y)
\ee
we  have 
\be\label{prod2}
\rho(X\times Y) = \rho(X) \chi(Y) +\chi(X) \rho(Y)
\ee
for all $X, Y$, which reduces to \eqref{prod1} for $X$ ($Y$) odd (even) dimensional, as required. Note, this same factorisation property is shared by the Ray-Singer and Franz-Reidmeister torsions \cite{RAY1971145}. 
Explicitly, 
\be
\rho(X\times Y) =\left\{
\begin{array}{lll}-(m_X+m_Y)\chi(X)\chi(Y), & \dim X=2m_X,& \dim Y=2m_Y\\ [5pt]\rho(X)\chi(Y), & X~\text{odd}, &Y~\text{even}\\[5pt]
\chi(X)\rho(Y), & X~\text{even},& Y~\text{odd}\\ [5pt]
0, & X~\text{odd},& Y~\text{odd}
\end{array}\right.
\ee

These relations are concisely expressed in terms of the Poincar\'e polynomials,
\be
P_X(z)=\sum_{k}^{d} z^k b_k(X).
\ee
Recall, the Euler characteristic is given by the Poincar\'e polynomial
evaluated at $z=-1$,
\be
\chi(X) = P_X(-1).
\ee
Similarly, \eqref{rho2} is simply expressed in terms of the derivative of the Poincar\'e polynomial,
\be
\rho(X) = P'_X(-1),
\ee 
where $P'=\partial_z P $. Then \eqref{prod2} follows immediately from $P_{X\times Y}(z)=P_{X}(z)P_{Y}(z)$,
\be
P'_{X\times Y}(-1)=P'_{X}(-1)P_{Y}(-1) + P_{X}(-1)P'_{Y}(-1). 
\ee

Using the freedom to add any amount of $\chi(X)$ to $\rho(X)$ while preserving \eqref{rho1} in odd dimensions allows us to introduce a one-parameter family of $\rho_\lambda$-characteristics
\be
\rho_\lambda(X) := \rho(X) + \lambda \chi(X).
\ee
Two notable members of the family are
\be
\begin{split}\label{rho3}
\rho_{\lfloor\tfrac{d}{2}\rfloor+1}(X) &= \sum_{p=0}^{d}(-1)^p (\lfloor\tfrac{d}{2}\rfloor+1-p) b_p(X)\\
\rho_{\tfrac{d}{2}}(X)& = \sum_{r}^{d}(-1)^r (\tfrac{d}{2}-r) b_r(X).
\end{split}
\ee
For  $\dim X =2m$ we have 
\be
\begin{split}
\rho_{\lfloor\tfrac{d}{2}\rfloor+1}(X)  &=  \chi(X) \\
\rho_{\tfrac{d}{2}}(X) &=  0. 
\end{split}
\ee
They also enjoy particularly  neat K\"unneth-type formulae,
\be
\rho_{\lfloor\tfrac{d}{2}\rfloor+1}(X\times Y) =\left\{
\begin{array}{lll}0, & X~\text{and}~ Y~\text{odd};\\ 
\rho_{\lfloor\tfrac{d}{2}\rfloor+1}(X)\rho_{\lfloor\tfrac{d}{2}\rfloor+1}(Y), & \text{otherwise};
\end{array}\right.
\ee
and 
\be
\rho_{\tfrac{d}{2}}(X\times Y) =\left\{
\begin{array}{lll}0, & X~\text{even},& Y~\text{even};\\ [5pt]\rho(X)\chi(Y), & X~\text{odd}, &Y~\text{even};\\[5pt]
\chi(X)\rho(Y), & X~\text{even},& Y~\text{odd};\\ [5pt]
0, & X~\text{odd},& Y~\text{odd}.
\end{array}\right.
\ee

Note, $\rho_\lambda$ cannot be defined  as a  sum of $I_p$, the number of $p$-simplices,   since the Euler characteristic is the unique (up to a proportionality constant) topological invariant that can be written as a (linear or non-linear) sum of $I_p$ \cite{Roberts:2002, YU2010794}. 
\subsection{Uniqueness}

Let us suppose there are  topological invariants $\alpha, \beta$ that (i) may be written as a linear combination of Betti numbers and (ii) obeys the K\"unneth type formula 
\be
\alpha(X \times Y)= \beta(X)\chi(Y),
\ee
for $X$ odd dimensional and $Y$ even dimensional. 
We show that $\kappa$ and $\lambda$ are proportional  to $\rho$. 

Let us first consider the original example of $d=11$  introduced in \cite{Duff:2010ss}.  Consider 4- and 7-manifolds $X^4$ and $X^7$ with Betti numbers
\be
a_p=b_p(X^4 \times X^7), \quad b_p=b_p(X^7), \quad c_p=b_p(X^4)
\ee
so that 
\be\label{k7}
a_r = \sum_{p+q=r} b_p c_q.
\ee

Let us suppose $\alpha(X^4 \times X^7)$and $\beta(X^7)$  are some linear combination of Betti numbers
 \be
 \alpha(X^4 \times X^7)=\sum_{p=0}^{5}A_pa_p, \qquad  \beta(X^7)=\sum_{p=0}^{3} B_pb_p.
 \ee
  If we now demand
\be
\alpha(X^4 \times X^7) =\beta(X^7)\chi(X^4)
\ee
we obtain     the multiplication table 
\be
\begin{array}{rrrrrrrrrr}
~&&B_0b_0& B_1b_1  &B_2b_2&B_3b_3\\
&&&&&&&\\
2c_0&&(A_0+A_4)c_0b_0&(A_1+A_5)c_0b_1&(A_2+A_5)c_0b_2&(A_3+A_4)c_0b_3\\
-2c_1&&(A_1+A_3)c_1b_0&(A_2+A_4)c_1b_1&(A_3+A_5)c_1b_2&(A_4+A_5)c_1b_3\\
c_2&&A_2c_2b_0&A_3c_2b_1&A_4c_2b_2&A_5c_2b_3&\\
\end{array}
\ee
where  we have used   \eqref{k7} and Poincar\'e duality to express all Betti numbers in terms of $c_p$ and  $b_q$ for $p<3, q<4$. 
We use the overall scaling to fix $A_0=11$. This implies the system of 12 equations,
\be
\begin{array}{ccccccccccccc}
2B_0&=&11+A_4&=&-A_1-A_3&=&2A_2\\[5pt]
2B_1&=&A_1+A_5&=&-A_2-A_4&=&2A_3\\[5pt]
2B_2&=&A_2+A_5&=&-A_3-A_5&=&2A_4\\[5pt]
2B_3&=&A_3+A_4&=&-A_4-A_5&=&2A_5
\end{array}
\ee
nine of which are independent, which (e.g.~sequentially eliminating $A_5, A_4\ldots$ first) immediately reduce   to 
\be
A_p = (-)^p (11-2p),\qquad B_p = (-)^p (7-2p).
\ee
Hence, 
\be
 \alpha(X^4 \times X^7)=\rho(X^4 \times X^7), \qquad \beta(X^7)=\rho(X^7)
 \ee
 Thus not only does $\rho$ satisfy the K\"unneth-type formula, it is the unique linear combination of Betti numbers to do so.

This immediately generalises to arbitrary $X^{d-d'}, X^{d'}$, assuming with out loss of generality $d'>d/2$.  Let $d=2n+1$ and $d'=2m+1$. 
Then we have a set of $n+m+1$ independent equations (again fixing $A_0=d$ ) yielding 
\be
A_p = (-)^p (d-2p),\qquad B_p = (-)^p (d'-2p).
\ee
for the $n+m+2$ coefficients $A_p, B_q$, $p=0,\dots n$, $q=0,\dots m$.

\subsection{Cosets}
Consider  a compact connected Lie group $G$ of rank $r$.  Its Poincar\'e polynomial is given by
\be
\label{poincare_G_and_H}
P_{G}(z)=\prod_{i=1}^{r} (1+z^{g_i}),\quad
\ee
where $g_i=2c_i-1$, with $c_i$ the degree of the $i^{th}$ Casimir invariant of $G$. 

Evidently,  the Euler characteristic vanishes
\be
\chi(G)=P_G(-1)=0.
\ee
 The same is true for $\rho$ if $r>1$,
 \be 
\rho(G)=P'_G(-1),
\ee
while 
\be
\rho(\Un(1))=1, \qquad \rho(\SU(2))=3.
\ee

For cosets on the other hand, both $\chi$ and $\rho$ can be non-trivial. Consider a compact connected Lie group $G$ of rank $r$ and a compact connected Lie subgroup $H$ of rank $s\leq r$. Let us denote their Poincar\'e polynomials  by
\be
\label{poincare_G_and_H}
P_{G}(z)=\prod_{i=1}^{r} (1+z^{g_i}),\quad
P_{H}(z)=\prod_{i=1}^{s} (1+z^{h_i})
\ee
where $g_i=2c(G)_i-1$, $h_i=2c(H)_i-1$, with  $c(G)_i$  and $c(H)_i$  the degree of the $i^{th}$ Casimir invariant of $G$ and $H$, respectively. Then
 the Poincar\'e polynomial of the coset manifold is 
\be\label{cosetP}
P_{G/H}(z)=\frac{\prod_{i=1}^s (1-z^{g_i+1})\prod_{i=s+1}^r (1+z^{g_i})}{\prod_{i=1}^s (1-z^{h_i+1})},
\ee
where 
\be
P_{\widehat{G/K}}(z) = \sum_{i=s}^{r} z^{g_i}
\ee
is the Poincar\'e polynomial of the Samelson subspace for the pair $(G, H)$. 
See, for example, \cite{Greub:1972v3}. 

From \eqref{cosetP} we have 
\be 
\begin{aligned}
\rho(G/H)=&\left(-\frac{\sum_{k=1}^s (1-z^{g_1+1})\cdots(g_k+1)z^{g_k}\cdots(1-z^{g_s+1})\ \prod_{i=s+1}^r (1+z^{g_i})}{\prod_{i=1}^s (1-z^{h_i+1})}\right.\\
&~~+\frac{\prod_{i=1}^s (1-z^{g_i+1})\ \sum_{k=s+1}^r(1+z^{g_{s+1}})...g_k z^{g_k-1}\cdots(1+z^{g_r})}{\prod_{i=1}^s (1-z^{h_i+1})}\\
&\left.~~ +\sum_{k=1}^s \frac{\prod_{i=1}^s (1-z^{g_i+1})\prod_{i=s+1}^r (1+z^{g_i}) (h_k+1)z^{h_k}}{(1-z^{h_1+1})\cdots(1-z^{h_k+1})^2\cdots(1-z^{h_s+1})} \right)\Biggr\rvert_{z=-1}
\end{aligned}
\ee
This goes as $\lim_{x\rightarrow 0}x^{r-(s+1)}$ and, hence,   $\rho$ can be  non-zero only if $r=s$ or $r=s+1$.

For $r=s$ we have  
\be 
\frac{2}{\dim G/H}\rho(G/H)=\chi(G/H)=-\prod_{i=1}^r\frac{g_i+1}{h_i+1},
\ee
whereas for $r=s+1$ we get
\be
\rho(G/H)=g_{s+1} \prod_{i=1}^s\frac{g_i+1}{h_i+1}.
\ee
A number of examples are given in \autoref{cosetrho}. 

Since  $G$ can be regarded as a bundle $H\rightarrow G\rightarrow G/H$  we have $\chi(G)=\chi(H)\chi(G/H)$, which  is trivially satisfied since  $\chi(G)=0$ for all $G$.
Nonetheless, although $\rho(G)=0$ for all $G$ with $r>1$, we can formally define, for $G$ and $H$ of equal rank
\be
\frac{\rho(G)}{\rho(H)}:=\lim_{z\rightarrow -1} \frac{P'_{G}(z)}{P'_{H}(z)}=\prod_{i=1}^r\frac{g_i}{h_i}.
\ee
Then
\be
\left.\frac{\rho(G)}{\rho(H)} \right|_{g_i\mapsto g_i+1, h_i\mapsto h_i+1} = -\chi(G/H),
\ee
resembling a K\"unneth-type formula. 
\begin{table}
\[
\begin{array}{|c|c|c|c|c|c|}
\hline\hline
M  & G & H & P_M(z) &\chi(M) & \rho(M) \\
\hline 
&&&&&\\
S^{2n}& \SO(2n+1) & \SO(2n) & 1+z^{2n} &2 & -2n \\[8pt]
S^{2n+1} & \SO(2n+2) & \SO(2n+1) & 1+z^{2n+1}&0 & 2n+1 \\[8pt]
\mathds{C}\mathds{P}^n& \SU(n+1) & \Un(n) & \sum_{k=0}^{n}z^{2k}&n+1 & -n(n+1) \\[8pt]
G/H & \SU(4) & \SU(2)\times \SU(2) & 1+z^4+z^5+z^9 & 0 & 10 \\[8pt]
  \hline\hline
\end{array}
\]
\caption{The $\chi$- and $\rho$-characteristics for various coset spaces $M=G/H$. }
\label{cosetrho}
\end{table}

\section{Mirror symmetry}\label{sec_mirror}
\FloatBarrier  
 \subsection{Conventional mirror symmetry: Calabi-Yau and Joyce}
 
Consider Calabi-Yau manifolds $X$ of complex dimension $n$. Denoting the Hodge numbers by $h^{p,q}=h^{n-p,n-q}$, the Betti numbers are
\be
b_k =\sum_ {p+q=k}h^{p,q}
\ee
and the Euler characteristic is 
\be
\chi(n):=\chi(X)=\sum_{k=0}^{k=n}(-1)^k b_k=\sum_{k=0}^{k=n}(-1)^k\sum_ {p+q=k}h^{p,q}.
\ee
Their mirrors $\check X$ are defined by 
\be\label{CYmirror}
\check h^{p,q} = h^{n-p,q},
\ee
where $\check h^{p,q} := h^{p,q}(\check X)$.

Then for odd complex dimension $n=2r+1$,
\be
\check \chi(2r+1)=-\chi(2r+1),
\ee 
where $\check \chi(n):=\chi(\check X)$
and
\be
\begin{split}
\text{IIA} ~\text{on} ~X &\equiv \text{IIB} ~\text{on}~ \check X \\
\text{IIB} ~\text{on} ~X &\equiv \text{IIA} ~\text{on}~ \check X
\end{split}
\ee
whereas for even complex dimension $n=2r$
\be
\check \chi(2r)=\chi(2r)
\ee 
and
\be
\begin{split}
\text{IIA} ~\text{on} ~X &\equiv \text{IIA} ~\text{on}~ \check X \\
\text{IIB} ~\text{on} ~X &\equiv \text{IIB} ~\text{on}~ \check X
\end{split}
\ee

Our focus will be on the Betti, rather than the Hodge,  numbers. It is therefore illustrative  to  rephrase the CY mirror symmetry in terms Betti numbers. The  CY3   Hodge  numbers are 
\be
h^{p,q} = 
\begin{pmatrix}
1 &0&0&1\\ 
0 &a&\check a&0\\ 
0 &\check a&a&0\\ 
1 &0&0&1\\ 
\end{pmatrix}
\ee
where $a=h^{1,1}$ and $\check a=h^{2,1}$ are the two independent entries. Written this way,
\be
(b_0,b_1,b_2,b_3)=(1,0,a,2+2\check a)
\ee
and 
\be
\chi(3)=2(a-\check a).
\ee
The mirror transformation is then 
\be
a\leftrightarrow \check a
\ee
which manifestly implies $\check \chi(3) = -\chi(3)$. In terms of the Betti numbers we have 
\be
\Delta b_2 = \check a -a = -\chi(3)/2, \qquad  \Delta b_3 = (2+2 a) -(2+2\check a)=\chi(3) 
\ee
as given in \autoref{Mirror}.

Similarly, for CY5 we have 
\be
h^{p,q} = 
\begin{pmatrix}
1 &0&0&0&0&1\\ 
0 &a&b&\check b& \check a&0\\ 
0 &b&c&\check c& \check b&0\\ 
0 &\check b&\check c& c&  b&0\\ 
0 &\check a&\check b& b&  a&0\\ 
1 &0&0&0&0&1\\ 
\end{pmatrix}
\ee
so that 
\be
(b_0,b_1,b_2,b_3,b_4,b_5)=(1,0,a,2b, 2\check b+c,2+2\check a+2\check c)
\ee
and
\be\label{chi5}
\chi(5)=2(a-\check a)-4(b-\check b)+2(c-\check c).
\ee
The mirror transformation is then 
\be
a\leftrightarrow \check a,\quad b \leftrightarrow \check b, \quad  c \leftrightarrow \check c,
\ee
which manifestly implies $\check \chi(5) = -\chi(5)$. 
Letting 
\be
A=a-\check a, \quad B=b-\check b, \quad C=c-\check c
\ee
in terms of the Betti numbers we have 
\be\label{betti5}
\Delta b_2 = -A, \qquad  \Delta b_3 = -2B, \qquad  \Delta b_4 = 2B-C, \qquad  \Delta b_5 = 2A+2C
\ee
as given in \autoref{Mirror}.

The extensions of the mirror map for Calabi-Yau manifolds \cite{Candelas:1989hd, Greene:1989cf, Greene:1990ud, Aspinwall:1990xe, Candelas:1990rm}, with holonomy $\SU(n)$, to  mirror maps for $d=7$ Joyce manifolds, with holonomy $G_2$, and $d=8$ $\text{Spin}(7)$ manifolds were considered in \cite{Shatashvili:1994zw, Acharya:1996fx,Joyce:1996g21,Joyce:1996g22}.  See also \cite{Braun:2019lnn}  for a recent treatment of mirror symmetry for generalised connected sum constructions of $G_2$ and $\text{Spin}(7)$  manifolds.  In these cases $b_1=0$. A further extension to more general manifolds, both even and odd  was considered in \cite{Duff:2010ss} without the condition that $b_1=0$. 

{A necessary condition for the quantum equivalence of type IIA/B  string theories compactified on mirror exceptional holonomy manifolds is that  the dimension of the space  of exactly marginal operators on the worldsheet is preserved \cite{Shatashvili:1994zw}. This is given by the number of moduli in NS-NS sector. Since the metric is determined by the associative 3-form for torsionless $G_2$ manifolds $X$, the number of geometric moduli is given $b_3(X)$ \cite{Joyce:1996g21,Joyce:1996g22}, so the number of NS-NS moduli is given by 
\be
b_2(X)+b_3(X).
\ee 
Similarly, for $\text{Spin}(7)$ manifolds $X$, the number of geometric moduli is given $b^{-}_{4}(X)+1$ \cite{Joyce:1996g21,Joyce:1996g22}, so the number of NS-NS moduli is given by 
\be
b_2(X)+b^{-}_{4}(X)+1.
\ee 
In all cases, a necessary condition for quantum equivalence is that the number of NS-NS moduli are preserved. Since $b_1=0$ is also preserved for supersymmetry,  this implies that the NS-NS sector is left invariant.  }

\subsection{Generalised mirror maps; $\chi \rightarrow 
\pm \chi $ in even $d$}

{The generalised mirror transformations are not peculiar to string theory or even supersymmetry, but given the preceding discussion we will nevertheless  maintain the property of leaving invariant the NS sectors and preserving/interchanging the RR sectors  of Type IIA  and IIB. Note, these conditions  are necessary but possibly not sufficient for quantum equivalence of type IIA/B compactified on mirror manifolds, hence the use of \emph{mirror maps} as opposed to \emph{mirror symmetry}. The known mirror symmetries are all examples of the generalised mirror maps considered here. }

Let $\chi_{\textsf{e}}(Y) $ be the sum of the even Betti numbers and $\chi_{\textsf{o}}(Y)$ the odd Betti numbers of the compactifying manifold $Y$,
 \be
 \chi_{\textsf{e}}(Y)=\sum_{k=0}b_{2k}(Y),\qquad 
  \chi_{\textsf{o}}(Y)= \sum_{k=0}b_{2k+1}(Y),
 \ee
so that 
\be
\chi(Y)= \chi_{\textsf{e}}(Y) -  \chi_{\textsf{o}}(Y)
\ee
To calculate the number of degrees of freedom, we note that for $\dim Y\leq 7$ preserving some supersymmetry, 
\be\label{d7dof}
f(\text{NS-NS})=f(\text{NS-R})=f(\text{R-NS})=f(\text{RR})=f/4
\ee
for both Type IIA and IIB, where $f(S)$ denotes the degrees of freedom of the sector $S$.  Moreover, from  \autoref{F}, using that a $p$-form in $D$ dimensions carries ${(D-2)!}/{(D-2-p)!p!}$ degrees of freedom,
we find
\be\label{RRdof}
f(\text{RR}) = 2^{(6-d)}\left(\chi_{\textsf{e}}(Y) + \chi_{\textsf{o}}(Y)\right)
\ee
and hence
\be
2^{(d-8)} f=\chi_{\textsf{e}}(Y) + \chi_{\textsf{o}}(Y)
\ee
for both Type IIA and Type IIB. This breaks down for $d=8$, where instead 
\be\label{d8chi}
\begin{split}
f_{\text{IIA}}(\text{RR}) &= \tfrac12\chi_{\textsf{o}}(Y)\\
f_{\text{IIB}}(\text{RR})  &= \tfrac12\chi_{\textsf{e}}(Y) \\
\end{split}
\ee
so that for all $d\leq 8$ we have the weaker statement
\be
f_{\text{IIA}}(\text{RR})+ f_{\text{IIB}}(\text{RR}) =  2^{(7-d)}\left(\chi_{\textsf{e}}(Y) + \chi_{\textsf{o}}(Y)\right).
\ee

For the generalized mirror symmetry we seek a transformation of the Betti numbers $\Delta b_n$ in $d=2m$ that preserves the number of degrees of freedom $\Delta f=0$ and for which $\Delta \chi=-2\chi$, $m$ odd, and $\Delta \chi=0$, $m$ even.   Taking into account \eqref{d7dof}, \eqref{RRdof} and \eqref{d8chi}, this is easily achieved by  demanding
 \be \label{mirrorAB}
 \chi_{\textsf{e/o}}(Y)   \mapsto  \left\{ \begin{array}{ll} \chi_{\textsf{o/e}}(Y) & m~~~{\text{odd}}; \\[5pt]
  \chi_{\textsf{e/o}}(Y) & m~~~{\text{even}}.
 \end{array}\right.
 \ee
 But $b_0=b_d$ is always unity and we do not want to change the number of supersymmetries. For $d<8$ this implies  $\Delta b_1 =0$, since $b_1$ and the degree of supersymmetry are bijective. This follows from the fact that any compact connected Ricci-flat manifold with $\dim X=n$ is of the form $X = (T^k\times X^{n-k})/\Gamma$, where $\Gamma$ is a freely acting finite subgroup of the isometries of $X^{n-k}$, where  $X^{n-k}$ is simply connected and $b_1(X^n)=k$ \cite{Fischer:1975aaa}. For $d=8$ the relation between $b_1$ and the number of supercharges breaks down. For example, $T^1\times X$ and $Y$, for $X$ a $G_2$-manifold and $Y$ a Calabi-Yau fourfold, both preserve $1/8$ supersymmetry but have distinct $b_1$. However, we shall none the less insist on $\Delta b_1 = 0$ in view of the  conventional mirror symmetry of Calabi-Yau manifolds. 
 
   Hence, we require 
\be \sum_{n=2}^{d-2}\Delta b_n=0\label{1e}\ee
\be \sum_{n=2}^{d-2} (-1)^n \Delta b_n =\left\{\begin{array}{ll}-2\chi & m~~\text{odd}\\ 0& m~~\text{even}\end{array}\right.\label{2e} \ee
and $d=6$ is the lowest dimension for which this mirror transformation is non-trivial. There \emph{are} non-trivial mirror symmetries for $d<6$, for example $T$-duality or mirror symmetries  taking $K3$ surfaces into themselves \cite{Aspinwall:1994rg}, but they act trivially on the Betti numbers  so are not captured by the mirror maps considered here. 

The $d=(6, 8, 10)$ transformations are shown in Table \ref{Mirror}. Note, for $d=6$ only $\Delta b_2, \Delta b_3$  are non-trivial so there are no free parameters in the mirror map. In $d=8,10$ there are one and two arbitrary integer parameters, respectively, defining a family of mirror maps. Further requirements may restrict these. For instance, in the case of Calabi-Yau manifolds the mirror symmetry (as opposed to map) is given by the finer Hodge number transformation \eqref{CYmirror} by the requirement of quantum equivalence of type II string theories compactified on the mirror manifolds.

\begin{table}[h!]
\center
$\begin{array}{|l|c|c|c|c|c|c|}
\hline
\hline
d&\Delta b_2&\Delta b_3&\Delta b_{4}&\Delta b_5&\Delta \chi&\Delta \rho\\
\hline
&&&&&&\\
6&-\chi/2&\chi&&&-2\chi&-2\rho\\
7&-\rho/2&\rho/2&&&0&-2\rho\\
8&n&0&-2n&&0&0\\
9&n&n&-2n&&0&0\\
10&-\chi/2-2B+C& -2B& 2B-C&\chi+4B& -2\chi&-2\rho\\
11&-\rho/2-2B+C&-\rho/2-4B+C&-C& \rho+6B-C&0&-2\rho\\
\hline
\hline
\end{array}$
\caption{Generalised mirror maps and their corresponding  $\Delta \chi, \Delta \rho$. By Poincar\'e duality $\Delta b_k=\Delta b_{d-k}$.  Here,  $n, A, B, C$ are  arbitrary integers   satisfying $ \chi=2A-4B+2C$ and $\rho=2A-4B+2C$, consistent with the special case of Calabi-Yau fivefolds, cf.~\eqref{chi5} and \eqref{betti5}. {Note, we demand $\Delta b^{+}_{d/2}=\Delta b^{-}_{d/2}$ consistent with the requirement that the NS-NS sector is left invariant for Calabi-Yau  and $\text{Spin}(7)$ compactifications.}}
\label{Mirror}
\end{table}

The degree of freedom  invariance under the mirror maps \eqref{mirrorAB}   can be made explicit by defining a formal Euler characteristic of R-R field strengths.     Since the NS-NS sector is  invariant, we focus on the bosonic R-R field strength content of Type IIA (even forms $F_0,F_2,F_4$) and Type IIB (odd forms $F_1,F_3,F_5^-$). Performing a Kaluza-Klein reduction on 
$
M=X \times Y
$, where $Y$ is a $d$-dimensional Riemannian and $X$ is a $(D=10-d)$-dimensional Lorentzian manifold, 
yields the $F$ content shown in Table \ref{F}.
 \begin{table}[h!]
\small{
$\begin{array}{lllllllllllllllllllllllllll}
F_0(X \times Y)&=&F_0(X)b_0(Y)\\
F_1(X \times Y)&=&F_0(X)b_1(Y)+F_1(X)b_0(Y)\\
F_2(X \times Y)&=&F_0(X)b_2(Y)+F_1(X)b_1(Y)+F_2(X)b_0(Y)\\
F_3(X \times Y)&=&F_0(X)b_3(Y)+F_1(X)b_2(Y)+F_2(X)b_1(Y)+F_3(X)b_0(Y)\\
F_4(X \times Y)&=&F_0(X)b_4(Y)+F_1(X)b_3(Y)+F_2(X)b_2(Y)+F_3(X)b_1(Y)+F_4(X)b_0(Y)\\
F_{5}^{-}(X \times Y)&=&F_0(X)b_5(Y)+ F_1(X)b_4(Y)+\cdots F_{m}^{+}(X)b_{(5-m)}^{-}(Y)+F_{m}^{-}(X)b_{(5-m)}^{+}(Y)...+F_5(X)b_0(Y)\\[5pt]
b_0(X \times Y)&=&b_0(X)b_0(Y)\\
b_1(X \times Y)&=&b_0(X)b_1(Y)+b_1(X)b_0(Y)\\
b_2(X \times Y)&=&b_0(X)b_2(Y)+b_1(X)b_1(Y)+b_2(X)b_0(Y)\\
b_3(X \times Y)&=&b_0(X)b_3(Y)+b_1(X)b_2(Y)+b_2(X)b_1(Y)+b_3(X)b_0(Y)+\\
b_4(X \times Y)&=&b_0(X)b_4(Y)+b_1(X)b_3(Y)+b_2(X)b_2(Y)+b_3(X)b_1(Y)+b_4(X)b_0(Y)\\
b_{5}^{-}(X \times Y)&=&b_0(X)b_5{}(Y)+b_1(X)b_4(Y)+ \cdots b_{m}^{+}(X)b_{(5-m)}^{-}(Y)+b_{m}^{-}(X)b_{(5-m)}^{+}(Y)+...+b_5(X)b_0(Y)\\
\end{array}$}
\caption{Comparison of  $F_p$ content of compactified IIA and IIB and the K\"unneth formula. The notation $F_p(X)b_q(Y)$ should read as there are $p$-form field strengths on $X$ with multiplicity $b_q(Y)$.}
\label{F}
\end{table}

Motivated by \autoref{F} let us introduce the formal $p$-form field strength
 ``Euler characteristic'' 
\be
 \chi(X)=F_0(X)-F_1(X)+F_2(X)+ \cdots
 \ee
and  ``Hirzebruch signature''
\be
\tau(X')=F_{m}^{+}(X) -F_{m}^{-}(X)~~m~~~\text{even}
\ee
akin to
\be
\chi(Y)=b_0(Y)-b_1(Y)+b_2(Y)+ \cdots
 \ee
 and
\be
\tau(Y)=b_{m}^{+}(Y) - b_{m}^{-}(Y)~~m~~ \text{odd}.
\ee
Note $X$ has Lorentzian spacetime signature and $\tau(X)$ is defined in 2 mod 4 while  $Y $ has Euclidean signature and $\tau(Y)$ is defined in 0 mod 4.

If we define
 \be
\Sigma^{\pm}_{\alpha}(M)=\frac{1}{2}[ \chi_\alpha(M) \pm \tau_\alpha(M)],\quad \alpha ={\sf e,o}
\ee
then field content of the R-R sector can be summarised by 
\be
\text{IIA}=\Sigma^{+}_{\sf e}(X)\Sigma^{-}_{\sf e}(Y)+\Sigma^{-}_{\sf e}(X)\Sigma^{+}_{\sf e}(Y)+\Sigma^{+}_{\sf o}(X)\Sigma^{-}_{\sf o}(Y)+\Sigma^{-}_{\sf o}(X)\Sigma^{+}_{\sf o}(Y)
\ee
\be
\text{IIB}=\Sigma^{+}_{\sf e}(X)\Sigma^{-}_{\sf o}(Y)+\Sigma^{-}_{\sf e}(X)\Sigma^{+}_{\sf o}(Y)+\Sigma^{+}_{\sf o}(X)\Sigma^{-}_{\sf e}(Y)+\Sigma^{-}_{\sf o}(X)\Sigma^{+}_{\sf e}(Y)
\ee
and
\[
\text{IIA}-\text{IIB}=[\chi(X) + \tau(X)][\chi(Y)-\tau(Y)]+[\chi(X) - \tau(X)][\chi(Y)+\tau(Y)]
\]
\be
=\chi(X)\chi(Y)+\tau(X)\tau(Y)=\chi(X\times Y)-\tau(X \times Y)
\ee
The virtue of this formalism is that invariance  under the generalized mirror symmetry \eqref{mirrorAB} is manifest
 \be
 \begin{array}{llllllllllll}
 \text{IIA}& \rightarrow &\text{IIB}&&        \text{IIB}& \rightarrow & \text{IIA}&&     m ~~\text{odd};\\[5pt]
  \text{IIA} &\rightarrow & \text{IIA} &&       \text{IIB}& \rightarrow &\text{IIB}&&     m ~~~\text{even}.
 \end{array}
 \ee
 for $\dim Y = 2m$. 
 
 However this formalism also reveals an unexpected symmetry, interesting in its own right, namely interchanging the Betti numbers and the fluxes while simultaneously  interchanging the spacetime and compact manifolds. See  \autoref{flux}.
 
\begin{table}[h!]
\footnotesize{
\[\begin{array}{|r|c|c|}
\hline
\hline
d&\text{Type IIA}&\text{Type IIB}\\
&&\\
\hline
0&(F_0+F_2+F_4)b_0&(F_1+F_3+F_5{}^-)b_0\\[4pt]
1&(F_0+F_1+F_2+F_3+F_4)b_0&(F_0+F_1+F_2+F_3+F_4)b_0\\[4pt]
2&(2F_0+2F_2+F_4)b_0+(F_1+F_3)b_1&2(F_1+F_3)b_0+(F_0+F_2+F_4/2)b_1/2+(F_0+F_2+F_4/2)b_1/2\\[4pt]
3&(F_0+F_1+F_2+F_3)(b_0+b_1)&(F_0+F_1+F_2+F_3)(b_0+b_1)\\[4pt]
4 &(F_0+F_2)(2b_0+b_2)+(2F_1+F_3)b_1&(F_1+F_3{}^-)(b_0+b_2{}^+) +(F_1+F_3{}^+)(b_0+b_2{}^-) + 2(F_0+F_2)b_1\\[4pt]
5 &(b_0+b_1+b_2)(F_0+F_1+F_2)&(b_0+b_1+b_2)(F_0+F_1+F_2)\\[4pt]
6  &(b_0+b_2)(2F_0+F_2)+(2b_1+b_3)F_1&(b_1+b_3/2)(F_0+F_2/2)+(b_1+b_3/2)(F_0+F_2/2) +2(b_0+b_2)F_1 \\[4pt]
7&(b_0+b_1+b_2+b_3)(F_0+F_1)&(b_0+b_1+b_2+b_3)(F_0+F_1)\\[4pt]
8&(2b_0+2b_2+b_4)F_0+(b_1+b_3)F_1&2(b_1+b_3)F_0+(b_0+b_2+b_4{}^+)F_1{}^-+(b_0+b_2+b_4{}^-)F_1{}^+\\[4pt]
9 &(b_0+b_1+b_2+b_3+b_4)F_0&(b_0+b_1+b_2+b_3+b_4)F_0\\[4pt]
10&(b_0+b_2+b_4)F_0&(b_1+b_3+b_5{}^-)F_0\\[4pt]
              \hline
\hline
\end{array}\]
\caption{Betti/flux (b/F ) duality  on $X^{(10-d)} \times Y^d$. }}
\label{flux}
\end{table}

Note, in deriving these results, we have assumed Poincar\'e duality
for manifold $Y$,
\be
b_p(Y)=b_{d-p}(Y),
\ee
and analogously we have freely dualised 
\be
F_p(X)\leftrightarrow F_{D-p}(X),
\ee
which is superficially possible in the R-R sector. 
As we shall see in  \autoref{sec_weyl}, a $p$-form and its dual can have different Weyl anomalies. Moreover in  \ref{Zduality} we shall see in greater detail how their partition functions may differ. However, since $Y$ is assumed to be Ricci flat and to preserve at least one supersymmetry, $X$ is a Minkowski spacetime and no duality anomaly can arise.  

Even if this were not the case,  one could  adopt the democratic formulation \cite{Bergshoeff:2001pv} whereby in $D=10$ spacetime dimensions IIA  includes not only $(F_0, F_2, F_4)$ but also their duals $(F_{10}, F_8, F_6)$ and IIB includes not only $(F_1,F_3,F_5{}^+)$ but also  $(F_9,F_7,F_5{}^{-})$.  Moreover, recognising that the anomalies given by the Seeley-deWitt coefficient vanish in odd dimensions and therefore after compactification to an even dimension the massive and massless contributions to the anomaly must cancel \cite{Duff:1982wm,Duff:1982gj}, as we demonstrate in \autoref{Z}. The assumption that IIA is duality anomaly free once its massive states are included is motivated by its M-theory uplift, where there can be no anomaly.  {Given   the equivalence between  type IIA and II B on a circle,  including the massive states in type IIB one can apply the same logic.}

\FloatBarrier
\subsection{Generalised mirror maps; $\rho \rightarrow \pm \rho$ in odd $d$}
Following the treatment of $\chi$ in even dimensions we let 
 \be
 \rho_{\textsf{e}}(Y)=-\sum_{k=0}2kb_{2k}(Y),\qquad 
  \rho_{\textsf{o}}(Y)= -\sum_{k=0}(2k+1)b_{2k+1}(Y),
 \ee
so that 
\be
\rho(Y)= \rho_{\textsf{e}}(Y) -  \rho_{\textsf{o}}(Y)
\ee
and the degrees of freedom $f$ for $d\leq 7$ are given by
\be
2^{(d-8)}d f= \rho_{\textsf{e}}(Y) + \rho_{\textsf{o}}(Y).
\ee

For the generalised mirror symmetry we seek a transformation of the Betti numbers $\Delta b_n$ in $d=2m+1$ that preserves the number of degrees of freedom $\Delta f=0$ and for which $\Delta \rho=-2\rho$, m odd, and $\Delta \rho=0$, $m$ even.   This is easily achieved by  
 \be \label{mirrorM}
 \rho_{\textsf{e/o}}(Y)   \mapsto  \left\{ \begin{array}{ll} \rho_{\textsf{o/e}}(Y) & m~~~{\text{odd}}; \\[5pt]
  \rho_{\textsf{e/o}}(Y) & m~~~{\text{even}}.
 \end{array}\right.
 \ee

Once again $b_0$ is always unity and we do not want to change the number of supersymmetries so cannot change $b_1$ in $d<8$ (and we apply this restriction for $d\geq8$).  Hence
\be \sum_{n=2}^{d-2}\Delta b_n=0\label{1e}\ee
\be \sum_{n=2}^{d-2} (-1)^n n  \Delta b_n  =\left\{\begin{array}{ll}-2\rho & m~~\text{odd}\\ 0& m~~\text{even}\end{array}\right. \label{3e}\ee
and  $d=7$ is the lowest dimension for which this mirror transformation is defined. The $d=(7, 9, 11)$ transformations are shown in  \autoref{Mirror}.
As a consistency check we note that for $d=7$ the condition \eqref{1e} reduces to
\be\label{b2b3}
\Delta b_2+\Delta b_3=0
 \ee
 as found in \cite{Shatashvili:1994zw,Joyce:1996g21,Joyce:1996g22} for $G_2$-manifolds. Note, $b_2+b_3$ is also the number of spin-1/2 fermions, cf.~\autoref{D=11}, and since we are preserving the degree of supersymmetry, this implies that  \eqref{b2b3} is also required by the invariance of the degrees of freedom.   Examples for $G_2$-manifolds may be found in  \autoref{examples}.

\begin{table}[h!]
\center
$\begin{array}{|c|c|c|c|c|c|}
\hline
\hline
(b_0,b_1,b_2,b_3)&k&(\Delta b_2, \Delta b_3)&\rho&f\\
\hline
&&&&\\
 (1,0,8,47) &0&(8,-8)&-16&224\\
(1,0,0,31 + 2k)&0,...,22,24,29,30&(12+k,-12-k)&-24-2k&128+8k\\
(1,0,1,30 + 2k)&0,...,19,21&(10+k,-10-k)&-20-2k&128+8k\\
(1,0,2,29 + 2k)&0,...,10,12,13,15&(8+k,-8-k)&-16-2k&128+8k\\
(1,0,3,28+2k)&0,..., 7,9,10&(6+k,-6-k)&-12-2k&128+8k\\
(1,0,4,27+2k)&0,1,2,3&(4+k,-4-k)&-8-2k&128+8k\\
(1,0,5,26)&0&(2,-2)&-4&128\\
\hline
\hline
\end{array}$
\caption{Generalised mirror map for examples of $G_2$ manifolds.}
\label{examples}
\end{table}

\section{Weyl anomalies}\label{sec_weyl}
\subsection{Weyl anomalies}

Weyl anomalies \cite{Deser:1976yx} take their most pristine form in the context of conformal field theories in a background gravitational field, for which the trace of the stress tensor is classically zero but nonzero at the quantum level (e.g.~conformal scalars and massless fermions in any $D$, Maxwell/Yang-Mills in $D =4$, $p$-form gauge fields in $D=2p+2$, Conformal Supergravity in $D=(2,4,6)$). For other
theories  (e.g.~Maxwell/Yang-Mills for $D \neq 4$, pure quantum gravity for $D > 2$, or any
theory with mass
terms) the ``anomalies'' will still survive, but will be accompanied by
contributions from
$ \langle g^{\alpha\beta} T_{\alpha\beta}\rangle$ expected anyway through the lack of
conformal invariance. Since 
the anomaly arises because the operations of regularizing and taking the trace
do not commute, the
anomaly ${ \cal A}$ in a theory which is not classically Weyl invariant may be defined as \cite{Duff:1977ay,Duff:1993wm,Casarin:2018odz}:
\begin{equation}
{\bf \cal A}^{\rm W}~:=g^{\alpha\beta} \langle T_{\alpha\beta} \rangle_{\tiny{\textrm{reg}}}-\langle g^{\alpha\beta}T_{\alpha\beta}\rangle_{\tiny{\textrm{reg}}}.
\label{A}
\end{equation}
Of course, the second term happens to vanish when the classical invariance is
present.  ${\bf \cal A}^{\rm W}$ is given by the Seeley-deWitt coefficient $B_d$ and will be local, which $g^{\alpha\beta} \langle T_{\alpha\beta}\rangle_{\tiny{\textrm{reg}}}$ in general is not.  That it still makes sense to talk of an anomaly in the absence of a symmetry is also familiar from the divergence of the axial 
vector current \cite{Kumura:1969wj,Delbourgo:1972xb} where the operations of regularizing and taking the divergence do not commute
\be
{\bf \cal A}^{\rm Axial}~:=\partial_{\mu}\langle(\sqrt{g}J^\mu{}_5)\rangle_{\tiny{\textrm{reg}}}-\langle\partial_{\mu}(\sqrt{g}J^\mu{}_5)\rangle_{\tiny{\textrm{reg}}}.
\ee
The anomaly can be understood as a quantum violation of the expectation value of a classical identity even if it is not forced to be identically zero by a symmetry, as is case for the axial vector current with massive fermions.

In $D=4$, for example, the fields in the massless sector of each theory will exhibit an on-shell\footnote {That is to say ignoring the Ricci terms.} trace anomaly \cite{Duff:1977ay,Duff:1993wm}
\begin{equation}
{\bf \cal A}^{\rm W}=A \frac{1}{32\pi^2}{}^*R{}^{\mu\nu\rho\sigma}{}^{*}R_{\mu\nu\rho\sigma}
\end{equation}
so that in Euclidean signature
\begin{equation}
\int_{X} d^4x \sqrt{g}{\cal A}^{\rm W}=A\chi(X)
\label{euler4}
\end{equation}
where $\chi(X)$ is the Euler characteristic of spacetime and $A$  is the anomaly  coefficient, which for conformal field theories is related to the central charges $c$ and $a$ by $A=720(c-a)$. See \autoref{D=11} for the anomaly coefficient contribution for each field.  

In particular,  $p$-form gauge fields $A_p$ in $D\neq 2p+2$ provide nice examples of theories that are scale invariant but not conformal invariant. In $D=4$  $A_p$  and their duals $A_{(2-p)}$ yield  \cite{Duff:1980qv}
\be
\int_{X} {\bf \cal A}^{\rm W}(A_2)-\int_X{\bf \cal A}^{\rm W}(A_0)=\chi(X),
\ee
and
\be
\int_X {\bf \cal A}^{\rm W}(A_3)=-2\chi(X).
\ee
These inequivalences were included in \autoref{D=11}. In arbitrary even dimensions
\be\label{tracediff}
\int_X {\bf \cal A}^{\rm W}(A_p)-\int_X{\bf \cal A}^{\rm W}(A_{\tilde p})=(-1)^p\frac{1}{2}(p- \tilde p)\chi(X).
\ee

 Such {\it quantum inequivalence} of $p$-forms and their duals has been called into question \cite{Siegel:1980ax,Grisaru:1984vk,Bern:2015xsa} on the grounds that their {\it total} stress tensors are the same and that the anomalous trace is unphysical.  Nevertheless, the Euler characteristic factors they provide in the partition functions are important for the subjects of  free energy \cite{Raj:2016zjp} and entanglement anomalies \cite{Donnelly:2016mlc}. This will the subject of \autoref{Z}. 
\FloatBarrier
\subsection{ $D=10$ Type IIA and $D=11$ M-theory: the roles of $\chi$ and $\rho$}

In this section, following \cite{Duff:2010ss,Duff:2010vy, Duff:2016whi}, we focus on compactifications of M-theory and its low-energy limit $D=11$ supergravity on seven-manifolds $X^7$ with Betti numbers $b_n=b_{7-n}$. See \autoref{D=11}. 
The compactifying manifolds we have in mind will be the product a Ricci flat manifold with special/exceptional holonomy and $b_1=0$, for example $G_2$-manifolds \cite{Papadopoulos:1995da, Joyce:1996g21, Joyce:1996g22}, and an $n$-torus with $b_1=n$. Such compactifications preserve at least one Poincar\'e supersymmetry \cite{Duff:1983vj,Duff:1985sf,Shatashvili:1994zw}
as opposed to one AdS supersymmetry resulting from a squashed $S^7$ \cite{Awada:1982pk}, for example, which has weak $G_2$ holonomy. The resulting field content is given in terms of the  Betti numbers  in  \autoref{D=11} in the case of $d=7$ dimensions.  For example, the moduli space of  torsion free $G_2$-structures is locally diffeomorphic to an open set of $H^3(X, \R)$ and $b_1=0$, so we recover the familiar result for $G_2$ compactifications, with a single $D=4$ gravitino and $b_3$ geometric moduli (plus $b_3$ moduli arising from the M-theory 3-form $A$). It holds similarly  for   $T^7$, $T^3\times K3$, $\text{CY}3\times S^1$, which covers all cases of concern. 

 Since
\be
\text{M} ~\text{on}~ (X^6 \times S^1) \equiv \text{IIA}~ \text{on} ~X^6
\ee
we can also read off the compactification of the Type IIA string  and its low-energy limit 
$D=10$ IIA supergravity on six-manifolds $X^6$ with Betti numbers $c_n=c_{6-n}$, where 
\be
(b_0, b_1, b_2, b_3)=(c_0,c_0+c_1,c_1+c_2, c_2+c_3)
\label{bc}
\ee
The number of degrees of freedom, $f$, and the anomaly coefficient $A$  for each massless supergravity field in $d=4$  \cite{Duff:1977ay, Duff:1993wm} is given in Table \ref{D=11}.  The total number of degrees of freedom is 
\be
f=4(b_0+b_1+b_2+b_3)=4(2c_0+2c_1+2c_2+c_3).
\ee

 We have seen in the previous subsection that the {\it  spacetime} Euler characteristic enters the anomaly calculation but more surprising perhaps is that the total coefficient $A$ can depend on the {\it internal} Euler characteristic:
\begin{equation}
A=-\frac{1}{24}\chi(X^6).
\label{euler6}
\end{equation}

If we now combine  \ref{euler4} and \ref{euler6} we are able to use the K\"unneth formula relating the Euler characteristics of  even-dimensional compact manifolds 
  \be
\chi(X)\chi(Y)=\chi (X \times Y)
\ee
to write\footnote{We stress that this is not the $d=10$ anomaly but that of the massless fields after compactification to $d=4$.}
\be
\int_X {\cal A}^{\rm W}=-\frac{1}{24}\chi(X^4)\chi(X^6)=-\frac{1}{24}\chi(X^4 \times X^6)
 \ee
 
Equally remarkable is that if we now repeat the calculation for $D=11$ supergravity on seven-manifolds $X^7$, the total
$A$ coefficient depends on the internal $\rho(X^7)$:
\begin{equation}
A=-\frac{1}{24}\rho(X^7).
\label{rho7}
\end{equation}

If we combine \ref{euler4} and \ref{rho7} and apply the odd dimensional analogue of the K\"unneth formula,
\be
\chi(X)\rho(Y)=\rho(X \times Y),
\ee
we find\footnote{Once again we stress that this is not the $d=11$ anomaly but that of the massless fields after compactification to $d=4$. Indeed, the $d=11$ anomaly vanishes since $d$ is odd.} 
\begin{equation}
\int_X {\cal A}^{\rm W}=-\frac{1}{24}\chi(X^4)\rho(X^7)=-\frac{1}{24}\rho(X^4 \times X^7).
\end{equation}

\subsection{Special case $X^d=X^{d-1} \times S^1$; $\rho(X^d)=\chi(X^{d-1})$}
 The $d=4$ anomalies from compactification of Type IIA on $X^6$  and those from compactification of M-theory on $X^7$  must be compatible with the duality
\be
\text{M} ~\text{on}~ (X^6 \times S^1) \equiv \text{IIA}~ \text{on} ~X^6.
\ee
This is indeed the case as may be seen from  \autoref{D=11}. The Betti numbers are related by  \ref{bc}
and hence
\be
\rho(X^{7})=\chi(X^6),
\ee
\be
\rho(X^4 \times X^{7})=\chi(X^4 \times X^6).
\ee
\FloatBarrier  
%

\section{Partition functions}\label{Z}

\subsection{Topological quantum field theory}

The $\rho$-characteristic naturally  appears in the log divergences of $p$-form partition functions $Z[A_0, A_1,\ldots]$. There is a long and rich history relating partition functions to topological invariants. See \cite{Birmingham:1991ty} for a review. Broadly speaking there are two classes of topological quantum field theories \cite{Birmingham:1988ap}: Schwarz-type and Witten-type. The Witten-type theories \cite{Witten:1988ze} have $Q$-exact actions, plus locally total derivative terms. Most directly relevant here, however,  are the Schwarz-type topological quantum field theories \cite{Schwarz:1978cn, Schwarz:1979ae}, which are defined to have  topological classical actions that are not total derivatives, Chern-Simons theory being the best known example.    Schwarz showed \cite{Schwarz:1978cn, Schwarz:1979ae} that a class of Schwarz-type $p$-form partition functions  yield the Ray-Singer torsion, which, as we previously observed, is formally of the same structure as $\rho$. Schwarz's perspective  was used in \cite{Nash:1992sf} to compute the Ray-Singer torsion of Lens spaces via Chern-Simons theory and  generalised to $p$-form theories with non-trivial cohomology   in \cite{Blau:1989bq}, where the metric independence  of the Ray-Singer torsion was re-derived  using the path integral and BRST-framework. 

Let us first review the Abelian Chern-Simons theory on a closed compact 3-manifold $X$,
\be
S_{\text{CS}}[A] = \int_M A\wedge dA.  
\ee
It is assumed, for now, that there are no zero-modes by considering $A$ valued in a flat $\pi_1(X)$-bundle $\mathcal{E}$ with trivial twisted Hodge-de Rham cohomology. 

Applying Schwarz's geometric   approach or the familiar BRST quantisation, see for example \cite{Blau:1989bq}, the free energy is given by 
\be
\begin{split}
F_{\text{CS}}[A] :=\ln Z_{\text{CS}}[A] &=-\frac{1}{4} \sum_{k=0}^{3} (-1)^k k\ln  \det \left( \frac{\Delta_k}{\mu^2} \right)\\
&= -\frac{1}{4} \sum_{k=0}^{3} (-1)^k k \ln \det \left(\Delta_k \right)+\frac{1}{2} \ln \mu \sum_{k=0}^{3} (-1)^k k \dim \Delta_k. 
\end{split}
\ee
Here $\mu$ is a dimensionful parameter that enters through the measure, for example  $DA=\prod_n \mu da_n$, where $A = \sum a_n A^\n$ for eigen-1-forms $A^\n$ with eigenvalues $\lambda_n$ of the Laplacian $\Delta_1$.

As we have assumed there are no zero-modes, we can straightforwardly  apply zeta-function regularisation of the partition function \cite{Hawking:1976ja} with no further subtleties. As  introduced by Ray and Singer \cite{RAY1971145} in the context of defining the Ray-Singer torsion, the zeta-regularised dimension and determinant of (twisted) Hodge-de Rahm Laplacians are defined by
\be
\dim\Delta_{k}|_{\text{reg}}:= \zeta_{k}(0), \quad \ln \det{} \Delta_k |_{\text{reg}} :=  -\zeta'_{k}(0),
\ee
where
\be
\zeta'_{k}(0) = \frac{d}{ds}\zeta_k(s)|_{s=0}, \qquad \zeta_k(s) = \sum_{n} \lambda_{n}^{-s}.
\ee
Since in odd dimensions $\zeta_k(0)=b_k$ we conclude
\be
\begin{split}
F_{\text{CS}}[A] &=-\frac{1}{2}\left(\ln T_{\text{RS}}(X, \mathcal{E}) + \rho(M) \ln \mu\right), 
\end{split}
\ee
where $T_{\text{RS}}(X, \mathcal{E})$ is the Ray-Singer torsion,
\be
T_{\text{RS}}(X, \mathcal{E}):= \frac{1}{2} \sum_{k=0}^{d} (-1)^k k \zeta'_{k}(0).
\ee
Since the we assume a trivial twisted de Rham-Hodge cohomology $T_{\text{RS}}(X, \mathcal{E})$ is metric independent \cite{RAY1971145}.

Assuming the classically topological action implies metric independence, the logarithmic term, required on dimensional grounds, must be controlled by a topological invariant. This is familiar from the zeta regularisation of not necessarily topological field theories in even dimensions, where it is the Euler characteristic that appears, cf.~\cite{Donnelly:2016mlc}. Of course, the Euler characteristic is trivial in odd dimensions and we see that the $\rho$-characteristic takes its place.

We can generalise this observation to arbitrary dimensions via Abelian $BF$ theory
\be\label{bf}
S_{BF}[A,B] = \int_M B\wedge dA,
\ee
where $B$ is a $p$-form and $A$ a $(d-p-1)$-form, possibly valued in a flat bundle $\mathcal{E}$. The action is invariant under the obvious local transformations,
\be
\delta B = d \Lambda_{{p-1}},\qquad \delta A = d \lambda_{{d-p-2}}
\ee
with redundancies $\delta \Lambda_{{p-1}} = d \Lambda_{{p-2}}, \ldots $ and $\delta \lambda_{{d-p-2}} = d \lambda_{{d-p-3}}, \ldots, \delta \lambda_{0} = 0$. 
The action is also invariant under the transformations
\be
\delta B = \Gamma_{p},\qquad \delta A =  \gamma_{{d-p-1}}
\ee
where $\Gamma, \gamma$ are harmonic forms, which can be used to trivialise the zero mode contributions to the partition function \cite{Blau:1989bq}. Note,  one can alternatively keep the harmonics modes in the partition function explicitly  \cite{Blau:1989bq}.

The Schwarz resolvent methodology or BRST quantisation \cite{Blau:1989bq} yields the free energy,
\be
F_{BF}[A,B] = -\tfrac{1}{2}\sum_{k=0}^{d} (-1)^k k \det \frac{\Delta_k}{\mu^2}-\tfrac{1}{4}\sum_{k=0}^{d} (-1)^k  \det \frac{\Delta_k}{\mu^2}. 
\ee
Applying zeta-regularisation we find,
\be
\begin{split}
F_{BF}[A,B]&= -T_{\text{RS}}(M, \mathcal{E})+ \ln \mu \sum_{k}^{d} (-1)^k k\zeta_k(0)\\
&~~~~~~-\tfrac{1}{4}\sum_{k=0}^{d} (-1)^k \zeta'_k(0)+\tfrac{1}{2}\ln\mu\sum_{k=0}^{d} (-1)^k \zeta_k(0).
 \end{split}
\ee
In odd dimensions, by Poincar\'e duality this reduces to,
\be
F_{BF}[A,B]= -T_{\text{RS}}(M, \mathcal{E})- \rho(M)\ln \mu. 
\ee
In even dimensions $d=2m$, we find 
\be
F_{BF}[A,B]=0,
\ee
where we have used $T_{\text{RS}}(M, \mathcal{E})=0$,  Poincar\'e duality and $\sum(-1)^k \zeta_k(0) = \sum (B_{d}^{(k)} -b_k)=0$, where $B_{i}^{(k)}$ are the integrated Seeley-DeWitt coefficients for $k$-forms (see \autoref{B4} for the $d=4$ case). 
\subsection{Duality anomalies}

\subsubsection{Electromagnetic duality and the $\chi, \rho$ characteristics}\label{Zduality}

The $\rho$-characteristic also appears in the context of duality anomalies of not necessarily  topological $p$-form field theories.  Recall, although classically a free $p$-form is equivalent to its dual $\tilde{p}$-form, where $\tilde{p}=d-2-p$, this correspondence  breaks down quantum mechanically \cite{Duff:1980qv}. In even dimensions there is a duality anomaly with consequences for physical properties,  such as the entanglement entropy \cite{Donnelly:2016mlc}. It is well-known that $\chi$, the Ray-Singer \cite{RAY1971145} and Reidmeister \cite{Reidemeister:1935vp} torsions play  important roles in the duality anomaly \cite{Schwarz:1984wk, Donnelly:2016mlc}. Here we discuss the appearance of $\rho$ in this same context.

Let us consider the free energy of an Abelian $p$-form field $A$ and its dual $\tilde{A}$, with classical actions
\be
S_{p}=\frac{1}{2e^2}\int_X F\wedge \star F,\qquad \tilde{S}_{\tilde{p}}=\frac{1}{2\tilde{e}^2}\int_X \tilde{F}\wedge \star \tilde{F},
\ee
where $F=dA +\mathcal{F}, \tilde{F}=d\tilde{A}+\tilde{\mathcal{F}}$, where $\mathcal{F}\in  2\pi H^{p+1}(X, \Z)$. Here $X$ is a closed Riemannian manifold. The electric $e$ and magnetic $\tilde{e}$ charges satisfy ${e}{\tilde{e}}=2\pi$ and have mass-dimensions
\be
[{e}]=p+1-d/2= \tfrac12({p}-\tilde{p}), \qquad [\tilde{e}]=-(p+1)+d/2= \tfrac12(\tilde{p}-p).
\ee
The electric potentials have mass-dimension $[A_{\sst(p)}]=p$, so 1-forms have the canonical geometric dimension, while the dual magnetic potentials have $[\tilde{A}_{\sst{(\tilde{p})}}]=\tilde{p}=d-p-2$. This is consistent with dimensionless couplings for  middle dimension field strengths. 

 The $p/\tilde{p}$  massless duality anomaly  is defined by 
\be
{\cal A}^{p,\tilde{p}} :=\ln \frac{{Z}_{p}}{Z_{\tilde p}}=F_{p} -F_{\tilde{p}}, 
\ee
where $F_{p} =\ln Z_{p}$ is the free-energy. 

We review here the derivation of the duality anomaly \cite{Donnelly:2016mlc}, which includes the non-zero mode contribution given in \cite{Schwarz:1984wk}. The partition function factors into three contributions,
\be
Z_{p} = Z^{\text{osc}}_{p}Z^{\text{zero}}_{p}Z^{\text{inst}}_{p},
\ee
corresponding to the oscillatory  modes, $A\in \Omega^{p}_{\text{exact}}(X)\oplus \Omega^{p}_{\text{coexact}}(X)$, the zero modes, $\Delta A=0$, and the instantons, $\mathcal{F}\in 2\pi H^{p+1}(X, \Z)$. 

\paragraph{Oscillatory modes}  Taking into account the ghost-for-ghosts \cite{Siegel:1980jj}, the partition function is formally given by
\be
Z_{p}^{\text{osc}} =\prod_{k=0}^{p} \left(\det{}^\prime \frac{\Delta_{p-k}}{\mu^2}\right)^{\tfrac{1}{2}(-1)^{k+1}(k+1)}=\prod_{k=0}^{p} \left(\det{}^\prime \frac{\Delta_{k}}{\mu^2}\right)^{\tfrac{1}{2}(-1)^{p+1-k}(p+1-k)} ,
\ee
where $\Delta_k$ is the Hodge-de Rahm Laplacian, $\det'$ denotes the restriction of the determinant to the subspace $\Omega^{p}_{\text{osc}}(X)\cong \Omega^{p}_{\text{exact}}(X)\oplus \Omega^{p}_{\text{coexact}}(X)$ of exact and coexact forms and $\mu$ is a mass-dimension-one constant required to make the path integral measure dimensionless. 

The free energy has  $\mu$-independent and  $\ln \mu$-dependent terms,
\be\label{freenon}
F^\text{osc}_{p} = \frac{(-1)^{}}{2}\sum_{k=0}^{p} (-1)^{p+1-k}(p+1-k)\left(\ln  \det{'}   \Delta_k -2\ln \mu^{\dim\Omega^{k}_{\text{osc}}(X)} \right).
\ee
  The $\dim\Omega^{k}_{\text{osc}}(X)$ and $ \det{'}   \Delta_k$ appearing in \eqref{freenon} can be evaluated using zeta-regularisation, 
\be
\dim\Omega^{k}_{\text{osc}}(X)|_{\text{reg}}:= \zeta_{k}(0), \quad \ln \det{}^\prime \Delta_k |_{\text{reg}}:=  \zeta'_{k}(0),
\ee
where
\be
\zeta'_{k}(0) = \frac{d}{ds}\zeta_k(s)|_{s=0}, \qquad \zeta_k(s) = \sum_{\lambda>0} \lambda^{-s}
\ee
and $\lambda$ are the non-zero eigenvalues of $\Delta_k$.
Thus 
\be
F^\text{osc}_{p} = -\tfrac{1}{2}\sum_{k=0}^{p}(-1)^{(p+1-k)}(p+1-k)\zeta_k'(0)+\ln(\mu)\sum_{k=0}^{p}(-1)^{(p+1-k)}(p+1-k)\zeta_k(0).
\ee

The duality anomaly in the oscillatory  mode sector is therefore given by $
{\cal A}^{p,\tilde{p}}_{\text{osc}}={\cal A}^{p,\tilde{p}}_{\text{osc}, \text{const}}+{\cal A}^{p,\tilde{p}}_{\text{osc}, \mu} $,
where 
\be\label{nonzeroanom}\begin{split}
{\cal A}^{p,\tilde{p}}_{\text{osc}, \text{const}} &= -\tfrac{1}{2}\sum_{k=0}^{d}(-1)^{(p+1-k)}(p+1-k)\zeta_k'(0), \\
 {\cal A}^{p,\tilde{p}}_{\text{osc}, \mu} &= \ln(\mu)\sum_{k=0}^{d}(-1)^{(p+1-k)}(p+1-k)\zeta_k(0),
 \end{split}
\ee
and  we have used Poincar\'e duality, $\zeta_k(s)=\zeta_{d-k}(s)$.

In even dimensions, $d=2m$, it is straightforward to demonstrate ${\cal A}^{p,\tilde{p}}_{\text{osc}} =0$. From \eqref{nonzeroanom} we have 
\be\begin{split}
{\cal A}^{p,\tilde{p}}_{\text{osc}, \text{const}}&= -(-1)^{p+1}(p+1-m)\tfrac{1}{2}\sum_{k=0}^{d}(-1)^{k}\zeta_k'(0), \\
 {\cal A}^{p,\tilde{p}}_{\text{osc}, \mu}  &= \ln(\mu) (-1)^{p+1}(p+1-m) \sum_{k=0}^{d}(-1)^{k}\zeta_k(0).
 \end{split}
\ee
Then the vanishing of ${\cal A}^{p,\tilde{p}}_{\text{osc}}$  immediately follows from 
\be
\sum_{k=0}^{d=2m}(-1)^{k}k\zeta_k(s)=0,
\ee
which is essentially a consequence of $\Omega^{p}_{\text{osc}}(X)\cong \Omega^{p}_{\text{exact}}(X)\oplus \Omega^{p}_{\text{coexact}}(X)$ and implies the vanishing of the Ray-Singer torsion,
\be
\ln T_{\text{RS}}(X)=\tfrac{1}{2}\sum_{k=0}^{d}(-1)^{k}k\zeta_k'(0)
\ee
 in even dimensions \cite{RAY1971145}. Note, due to the non-trivial cohomology $T_{\text{RS}}(X)$ is metric dependent. This can be rectified by defining $\hat{T}_{\text{RS}}(X):=V_{H^\bullet(X)} T_{\text{RS}}(X)$, where $V_{H^\bullet(X)}$ is an element of the cohomology determinant line, $\det H^\bullet(X):=\bigotimes_{k=0}^{d}(\det H^k(X))^{(-1)^k}$, given by an arbitrary orthonormal basis in the cohomology with respect to the inner product induced by the representation of $H^\bullet(X)$ by harmonic forms and the Hodge $p$-form inner product on $\Omega^\bullet(X)$. For a nice introduction to these notions, see the proof of the Cheeger-M\"uller theorem \cite{Muller:1978, Cheeger:1979}, using the Witten deformation of the Laplacian \cite{Witten:1982im},  given in \cite{Braverman:2003vn}.

From \eqref{nonzeroanom}, for $d$ odd we have,
\be\label{anonodd}\begin{split}
{\cal A}^{p,\tilde{p}}_{\text{osc}, \text{const}}&= (-1)^{p+1}\tfrac{1}{2}\sum_{k=0}^{d}(-1)^{k}k\zeta_k'(0)=(-1)^{p+1}\ln  T_{\text{RS}}(X), \\
{\cal A}^{p,\tilde{p}}_{\text{osc}, \mu} &= \ln(\mu) (-1)^{p} \sum_{k=0}^{d}(-1)^{k}k\zeta_k(0)= (-1)^{p+1}\ln(\mu)  \rho(X).
 \end{split}
\ee
We see that $\rho$ characterises the logarithmic contribution to the odd-dimensional oscillatory mode anomaly.

\paragraph{Zero modes} The action vanishes for zero-modes, so their contribution to the partition function is given solely by the volume of the space of flat connections modulo large gauge transformations, $A\sim A+2\pi \theta$, where $\theta\in H^p(X, \Z)$. The ghost $k<p$ forms contribute similarly. This gives

\be
Z_{p}^{\text{zero}} =\prod_{k=0}^{p} \det \left(\frac{2\pi}{e^2} \mu^{2(k+1)}\Gamma_{p-k}\right)^{\tfrac{1}{2}(-1)^{k}}=\prod_{k=0}^{p} \det \left(\frac{2\pi}{e^2}  \mu^{2(p-k+1)}\Gamma_{k}\right)^{-\tfrac{1}{2}(-1)^{p+1-k}},
\ee
where $\Gamma_{k}$ is the $b_k(X)\times b_k(X)$ Jacobian matrix corresponding to the change from the path integral measure basis to the topological basis of  $\text{Free} H^k(X, \Z)$. 

Hence,
\be
F_{p}^{\text{zero}} = -\ln( \mu)\sum_{k=0}^{p}(-1)^{p+1-k}(p+1-k)b_k  -\frac{1}{2} \sum_{k=0}^{p}(-1)^{p+1-k}\ln \det \left(\frac{2\pi}{e^2}  \Gamma_{k}\right)
\ee
and 
\be\label{azero}\begin{split}
{\cal A}^{p,\tilde{p}}_{\text{zero}, \text{const}}&= (-1)^{p}\tfrac{1}{2}\sum_{k=0}^{d}(-1)^{k}\ln \det \left(\frac{2\pi}{e^2} \Gamma_{k}\right)+\tfrac{1}{2} \ln \det \left(\frac{2\pi}{e^2} \Gamma_{p+1}\right)\\
&= (-1)^{p}\tfrac{1}{2}\ln \left(\frac{2\pi}{e^2} \right)\chi(X)+(-1)^{p}\ln T_\text{R}(X)+\tfrac{1}{2} \ln \det \left(\frac{2\pi}{e^2} \Gamma_{p+1}\right),\\
{\cal A}^{p,\tilde{p}}_{\text{zero}, \mu} & =\ln(\mu)\times  \left\{ \begin{array}{cc} \tfrac{(-1)^{p}}{2}(\tilde p-p) \chi(X) &\text{even}\\[5pt](-1)^{p} \rho(X) &\text{odd} \end{array} \right.
 \end{split}
\ee
where we have used $(\frac{2\pi}{e^2}  \Gamma_{k})^{-1}=\frac{2\pi}{\tilde{e}^2}  S\Gamma_{d-k}$ for some unit determinant  matrix $S$, and introduced the Reidemeister torsion,
\be
\ln T_\text{R}(X) = \tfrac{1}{2}\sum_{k=0}^{d}(-1)^{k}\ln \det \left(\Gamma_{k}\right).
\ee

\paragraph{Instantons} The instanton contribution is given by the sum over field strengths belonging to $2\pi H^{p+1}(m, \Z)$, which we assume to be torsion free $H^{k}(m, \Z)\cong \Z^{b_k}$ so that
\be
Z^{\text{inst}}_{p}=\sum_{F\in \Z^{b_{p+1}}} e^{-S[F]}.
\ee
In appendix A of \cite{Donnelly:2016mlc} it is shown that 
\be
Z^{\text{inst}}_{\tilde{p}} =  \det \left(\frac{2\pi}{e^2} \Gamma_{p+1}\right)^{1/2} Z^{\text{inst}}_{p}
\ee
so
\be
{\cal A}^{p,\tilde{p}}_{\text{inst}} = -\tfrac{1}{2}\ln \det \left(\frac{2\pi}{e^2} \Gamma_{p+1}\right).
\ee

\paragraph{Total}
To summarise, the various contributions to the duality anomaly are given in \autoref{dualityanom}.
\begin{table}[h!]
\small
\[
\begin{array}{|c|c|c|c|c|c|c|cccccc}
\hline\hline
\text{Anomaly} & \text{Massless}&\text{Massless}& \text{Massive}& \text{Massive}\\ [8pt]
 & d=2n& d=2n+1&d=2n& d=2n+1\\ [8pt]

\hline\hline
			&&&&\\

\text{Oscillatory}	&0 &	(-1)^{p+1}\ln  T_{\text{RS}}(X)	&\tfrac{(-1)^p}{2} \chi(M)\ln (\frac{\mu^2}{m^2
})&0\\	[8pt]
			& &+	 (-1)^{p+1}\rho(X) \ln \mu&&	\\	[8pt]
			\hline
			&&&&\\
{\text{Zero}}  	& (-1)^{p}\ln \left(\frac{e}{\tilde{e}} \right)^{\tfrac{\chi(X)}{2}}+\tfrac{1}{2} \ln \det \left(\frac{2\pi}{e^2} \Gamma_{p+1}\right) & (-1)^{p}\ln T_\text{R}(X)+\tfrac{1}{2} \ln \det \left(\frac{2\pi}{e^2} \Gamma_{p+1}\right) &0&0 \\[8pt]
			& + \tfrac{(-1)^{p}}{2}(\tilde p-p) \chi(X) \ln \mu&	 + (-1)^{p} \rho(X)	 \ln \mu&&	\\	[8pt]
\hline
			&&&&\\

{\text{Instantons}}& -\tfrac{1}{2}\ln \det \left(\frac{2\pi}{e^2} \Gamma_{p+1}\right)	& -\tfrac{1}{2}\ln \det \left(\frac{2\pi}{e^2} \Gamma_{p+1}\right) &0&0	\\	[8pt]
	& &	&&	\\
	\hline	
				&&&&\\

		\text{Total} & \tfrac{(-1)^{p}}{2} \chi(X) \ln \frac{e\mu^{ \tilde p}}{\tilde{e}\mu^{{p}}}&0 
 		&\tfrac{(-1)^p}{2} \chi(M)\ln (\frac{\mu^2}{m^2
})&0\\[8pt]
			& &	&&	\\	[8pt]
		\hline
		\hline

\end{array}
\]
\caption{The duality anomaly contributions for massless and massive $p$-forms.}
\label{dualityanom} 
\end{table}

Note, the cancellation of the anomaly in odd dimensions makes use of the Cheeger-M\"uller theorem, which for torsion free cohomology reads
\be
\ln T_{\text{RS}}(X, \mathcal{E}) = \ln T_{\text{R}}(X, \mathcal{E}). 
\ee
In the case that there are no zero modes, the $\rho$-characteristic determines the  $\ln \mu$ contribution to the anomaly.

Observe that the zero-mode contribution $\tfrac{(-1)^{p}}{2}(\tilde p-p) \chi(X) \ln \mu$ corresponds to the difference in the trace anomalies for $p$- and $\tilde p$-forms \cite{Duff:1980qv}, cf.~\eqref{tracediff}.  For even spheres, this agrees with the results of \cite{Raj:2016zjp}.

\paragraph{Torsion groups} Throughout we assumed the cohomology was torsion free. Relaxing this condition, $H^{k}(m, \Z)\cong \Z^{b_k} \oplus T^{k}$,  the instanton partition function picks up a factor of 
$|T^{p+1}|$, while the zero-modes contribute $\prod_{k=0}^{p}|T^{k}|^{(-1)^{p+1-k}}$ \cite{Donnelly:2016mlc}. Hence, the anomaly due to the torsion groups is given by 
\be
\begin{split}
A_{\text{tor}}^{p, \tilde{p}} &=(-1)^{p+1}\sum_{k=0}^{p+1}(-1)^{k}\ln |T^k|- (-1)^{\tilde{p}+1}\sum_{k=0}^{\tilde{p}}(-1)^{k}\ln |T^k|\\&= (-1)^{p+1}\sum_{k=0}^{d}(-1)^{k}  \ln |T^k|.
\end{split}
\ee
In even dimensions this is identically zero. In odd dimensions, it cancels against the difference between the  Ray-Singer and Reidemeister torsions  in the presence of torsion in the cohomology \cite{Cheeger:1979}
\be
\ln T_{\text{RS}}(X) - \ln T_{\text{R}}(X) = \sum_{k=0}^{d}(-1)^{k}  \ln |T^k|.
\ee

\subsubsection{Massive duality anomalies and Kaluza-Klein compactification}

Here we consider the vanishing massless  $p=0, \tilde{p}=\hat{d}-2$ duality anomaly in  $\dim \hat X= \hat{d}=d+1=2n+1$ dimensions and its Kaluza-Klein reduction on $\hat X=X \times S^1, \dim X=2n $.   In $\hat{d}$ dimensions we have,
\be\label{d503}
S_0= \frac{1}{2\hat{e}^2}\int_{\hat X} \hat{F}_{\1}\wedge \star \hat{F}_\1,\qquad \tilde{S}_{d-1}= \frac{1}{2\hat{\tilde{e}}^2}\int_{\hat X} \hat{\tilde{F}}_{\sst{(d)}}\wedge \star\hat{ \tilde{F}}_\sst{(d)}
\ee
where  the electric/magnetic charges satisfy $\hat{e}\hat{\tilde{e}}=2\pi$.
Compactifying  on $\hat X = X\times S^1$, where $S^1$ has radius $R$, we have
\be\label{d40}
S_0= \frac{1}{2e^2}\int_{X} F_{\1}\wedge \star F_\1 + \frac{1}{2e'^2}\int_{X} {F}_{\0}\wedge \star {F}_\0 + \sum_{n=-\infty}^{\infty} S_{0}^{\n}
\ee
where $S_{0}^{\n}\sim \frac{1}{2e^2}\int ( F^\n_{\1}\wedge \star F^\n_\1 + \frac{n^2}{2R^2} (A_\0^\n)^2)$ correspond to the massive Kaluza-Klein scalars and 
\be
e= \frac{\hat{e}}{\sqrt{2\pi  R}},\qquad e' = \hat{e}\sqrt{2\pi R}.
\ee
 Similarly, 
\be\label{d43}
\tilde{S}_{\sst{(d-1)}}= \frac{1}{2{\tilde{e}'}{}^2}\int_{X} \tilde{F}_{\sst{(d)}}\wedge \star \tilde{F}_\sst{(d)} + \frac{1}{2{\tilde{e}}{}^2}\int_{X} \tilde{F}_{\sst{(d-1)}}\wedge \star \tilde{F}_\sst{(d-1)} + \sum_{n=-\infty}^{\infty} S_{\sst{(d-1)}}^{\n}
\ee
where 
\be
{\tilde{e}'}= \frac{\hat{{\tilde{e}}}}{\sqrt{2\pi R}},\qquad {\tilde{e}} = \hat{{\tilde{e}}}\sqrt{2\pi R}
\ee
so that $e\tilde{e}=2\pi$ and ${e'}\tilde{e}'=2\pi$. 

From the $d$-dimensional perspective the massless contribution to the duality anomaly is 
\be
\frac{\chi(M)}{2}\left[\ln \left(\frac{e}{\tilde e}\right)-\ln \left(\frac{{e'}}{\tilde e'}\right)\right] +\chi(M)\ln \left(\mu\right)-2\chi(M)\ln \left(\mu\right)=-\chi(M)\ln (2\pi R\mu).
\ee

Since the $\hat{d}$-dimensional  anomaly vanishes this must be cancelled by the duality anomaly of the infinite tower of massive Kaluza-Klein modes. The duality anomaly of massive $p$-forms was treated in \cite{Kuzenko:2020zad}. In this case a $p$-form is classically dual to a $(d-p-1)$-form of the same mass, so that the difference for each $n$ of the massive 0-forms and $(d-1)$-forms contributions in \eqref{d40} and \eqref{d43} to the free energy is the correct duality anomaly.  In this case there are only the non-zero modes to contribute. In $d$-dimensions the $p, \tilde{p}=d-p-1$ duality anomaly with mass $m$ is given by,
\be\label{massA}
{\cal A}^{p,\tilde{p}}_{m} = (-1)^p \frac{1}{2}\sum_{k=0}^{d}(-1)^k \ln \det \left(\frac{\Delta_k +m^2}{\mu^2}\right).
\ee
Note, unlike the massless case there is no factor of $k$ in the sum - it is of `Euler form' as opposed to `Ray-Singer form' (of course, this distinction is only relevant in odd dimensions). Consequently, the anomaly vanishes in odd dimensions, due to Poincar\'e duality, but is non-zero in even dimensions, just as in the massless case. In the massless case the non-zero mode contribution to the anomaly in even dimensions vanished, whereas in the massive case it is non-vanishing and is the only contribution - there is nothing else for it to cancel against. 

The determinant in \eqref{massA} can be regulated using the spectral Hurwitz function.  For a Laplace-type operator $A$ with non-negative spectrum $\{\lambda\}$ the spectral Hurwitz function is defined by the meromorphic continuation of  
\be\label{HZ}
\zeta_A(s, a) := \sum_{\lambda>0}(\lambda+a)^{-s},\qquad a>0.
\ee
Note, since the spectrum of $\Delta_k$ is non-negative we must include its zero-modes in the extended spectral Hurwitz function,
\be\label{exHZ}
Z_k(s, a) := \sum_{\lambda\in\text{spec}(\Delta_k)}(\lambda+a)^{-s}= b_k a^{-s} + \zeta_k(s, a).
\ee
Then 
\be
Z_k(0, a)= b_k   + \zeta_k(0, a)
\ee
and
\be
Z'_k(s, a)= -b_k a^{-s} \ln a  + \zeta'_k(s,a)
\ee
so that 
\be
Z'_k(0, a)= -b_k  \ln a  + \zeta'_k(0,a).
\ee

We shall use $Z_k(s, a)$ to regulate (letting $a=m^2$) the determinant and dimension of $\Delta_k +a$:
\be
Z_{k}(0, a)=:\dim \left({\Delta_k + a}\right)|_{\text{reg}},\qquad Z_{k}'(0, a)=: -\ln \det \left({\Delta_k + a}\right)|_{\text{reg}}
\ee
so that 
\be\label{massA2}
\begin{split}
2(-1)^p{\cal A}^{p,\tilde{p}}_{m}|_{\text{reg}} &= \sum_{k=0}^{d}(-1)^k \ln \det \left({\Delta_k +a}\right)|_{\text{reg}}- \ln\left(\mu^2\right) \sum_{k=0}^{d}(-1)^k  \dim \left({\Delta_k +a}\right)|_{\text{reg}}\\
&= -\sum_{k=0}^{d}(-1)^k Z'_k(0, a)- \ln \mu^2 \sum_{k=0}^{d}(-1)^k Z_k(0, a)\\
&=\chi(M)\ln a - \sum_{k=0}^{d}(-1)^k \zeta'_k(0, a)-\chi(M) \ln \mu^2 - \ln \mu^2 \sum_{k=0}^{d}(-1)^k \zeta_k(0, a)\\
&=\chi(M)\ln\left( \frac{a}{\mu^2}\right) - \sum_{k=0}^{d}(-1)^k \zeta'_k(0, a) - \ln \mu^2 \sum_{k=0}^{d}(-1)^k \zeta_k(0, a)
\end{split}
\ee

Noting that $\sum_{k=0}^{d}(-1)^k \zeta_k(s, a)=0$, cf. \cite{Buchbinder:2008jf}, we conclude
\be
2(-1)^p {\cal A}^{p,\tilde{p}}_{m}|_{\text{reg}} =- \chi(M)\ln (\frac{\mu^2}{m
^2}).
\ee
Hence, we obtain a non-zero massive $p$-form duality anomaly in even dimensions consistent with \cite{Kuzenko:2020zad}, but in contrast to \cite{Buchbinder:2008jf}. The difference is due to our use of  \eqref{exHZ}, as opposed to \eqref{HZ} applied in \cite{Buchbinder:2008jf}, for the regularisation. Its non-vanishing is required on physical grounds to cancel the massless anomaly in the Kaluza-Klein reduction, as we  illustrate below.

The total anomaly for the $p=0, \tilde{p}=d-1$ Kaluza-Klein tower with masses $m_n=|n|/R$ is then
\be
 {\cal A}^{p,\tilde{p}}_{\text{Kaluza-Klein}} =- \chi(M) \sum_{n=1}^{\infty}\ln (\frac{\mu R}{n}) =- \chi(M) \ln ({\mu R}) \sum_{n\in \Z_{\ne 0}} \frac{1}{|n|^{0}}+ \chi(M) \sum_{n\in \Z_{\ne 0}}\ln |n|.
\ee
Let us zeta-regulate the infinite sums,
\be
\left.\sum_{n\in \Z_{\ne 0}} \frac{1}{|n|^{0}}\right|_{\text{reg}}=2\zeta(0)=-1\qquad \left.\sum_{n\in \Z_{\ne 0}}\ln |n|\right|_{\text{reg}}= -2\zeta'(0)=2\ln \sqrt{2\pi}.
\ee
Hence,
\be
 {\cal A}^{p,\tilde{p}}_{\text{Kaluza-Klein}} = \chi(M) \ln ({\mu R})+2 \chi(M)\ln \sqrt{2 \pi}= \chi(M) \ln ({2\pi \mu R})
\ee
and the massless $d$-dimensional  duality anomaly is cancelled precisely by the duality anomalies of the massive Kaluza-Klein tower.

Of course, the cancellation between the massless and massive duality anomalies witnessed above had to be the case; the  $\hat{d}$-dimensional  anomaly is always vanishing. 

If one insists that type IIA supergravity is duality anomaly free, then the accompanying massive tower of states is essential. This provides yet another rational for the inclusion of D-branes. Equivalently, it provides yet another rational for M-theory. The $D=11$ theory is duality anomaly free and so its compactification to type IIA must be duality anomaly free. Duality anomaly freedom is ensured by M-theory.
The type IIA R-R 1-form potential derives from the metric, so really this argument assumes duality anomaly freedom for gravity. {Since gravity is non-Abelian, unlike the R-R sector, it is not  clear that gravitational duality can be implemented. At least in the free case, however, classically dual graviton theories are well-known, cf.~for example  \cite{Hull:2001iu, deMedeiros:2002qpr, deMedeiros:2003osq},  and one can at least test the duality anomaly freedom for the free theory.  Note, from the point of view of $E_{11}$ \cite{West:2001as} there should   be, and there is \cite{Glennon:2020uov}, an interacting  dual graviton theory and  consistency of M-theory would imply duality anomaly freedom.}

\section{Conclusions}

In this paper the topological invariant 
\be\label{rho0}
\rho(X) = 7 b_0 - 5 b_1 +3 b_2 -b_3,
\ee
 introduced in \cite{Duff:2010ss, Duff:2010vy, Duff:2016whi} to describe the compactification of M-theory from  $D=11$ to $D=4$,  is generalized to arbitrary odd dimensions $\dim X = 2m+1$,
\be\label{rho1}
\rho(X) = \sum_{p=0}^{m} (-1)^p(\dim X-2p) b_p(X) ,
\ee
and shown in several respects to behave as an odd-dimensional analogue of the Euler characteristic in even dimensions. For example it obeys the  K\"unneth-type formula,
\be\label{prod1}
\rho(X \times Y) = \rho(X)\chi(Y),
\ee
where $X$ and $Y$ are even and odd dimensional, respectively, and is the unique linear combination of betti numbers to do so.

Whereas $\chi$ is related to the Poincare polynomial $P(z)$ by $P(-1)$,  $\rho$ is given by  $P'(-1)$. Both  vanish for group manifolds, except for $\rho(\Un(1))=1$ and $\rho(\SU(2))=3$. For cosets $G/H$ on the other hand 
\be 
\chi(G/H)=-\prod_{i=1}^r\frac{g_i+1}{h_i+1},
\ee
when $r:=\text{rank} ~G = s:= \text{rank}~ H$, and
\be
\rho(G/H)=g_{s+1} \prod_{i=1}^s\frac{g_i+1}{h_i+1}.
\ee
when $r=s+1$,
where
 $g_i=2c(G)_i-1$, $h_i=2c(H)_i-1$, with  $c(G)_i$  and $c(H)_i$  the degree of the $i^{th}$ Casimir invariant of $G$ and $H$, respectively. For example $ \chi(S^{2n})=2,\rho(S^{2n})=2n;
 \chi(S^{2n+1})=0,\rho(S^{2n+1})=2n+1$.

The $\rho$-characteristic also makes an appearance in the $d=4$ Weyl anomaly of the massless sector of  M-theory compactifications.   For $d=11$  M-theory on $X^4(spacetime) \times Y^7(internal)$  the $d=4$ on-shell Weyl anomaly ${\cal A}^{\rm W}$ is given by
 \be
\int {\cal A}^{\rm W}=-\frac{1}{24}\chi(X^4)\rho(Y^7)=-\frac{1}{24}\rho(X^4 \times Y^7)
\label{rho7}
\ee
 on using the K\"unneth rule (\ref{kunrho}) and hence vanishes when $Y^7$ (and therefore $X^{11}$) is self-mirror. By virtue of the special case  (\ref{special}), this is entirely consistent with the equivalence
 \be
 \text{M-theory}~ \text{on}~ X^{10} \times S^1 \equiv \text{IIA}~\text{on} ~X^{10}.
 \ee

  Odd dimensional generalized mirror maps are characterised by $\rho$, {where $\rho \mapsto (-)^{m}\rho$ for $d=2m+1$, in analogy to $\chi$ for even dimensions. If $\rho=0$ the manifold is `self-mirror'  and for the case of $G_2$ manifolds this defines an axis of symmetry \cite{Joyce:1996g21,Joyce:1996g22, Duff:2010ss}}. Conventional mirror symmetry relates  theories with Ricci-flat manifolds of special holonomy $X$, for which  $b_1=0$, for example Calabi-Yau, $G_2$, $\text{Spin}(7)$. Generalized mirror symmetry permits non-vanishing $b_1$, for example $(X \times T^k)/\Gamma$ with $b_1=k$, which in fact exhaust all closed Ricci-flat manifolds preserving some supersymmetry.

The $\rho$-characteristic naturally  appears in the log divergences of $p$-form partition functions $Z[A_0, A_1,\ldots]$ in odd dimensions.  We find
\be
\begin{split}
F_{\text{CS}}[A] &=-\frac{1}{2}\left(\ln T_{\text{RS}}(M, \mathcal{E}) + \rho(M) \ln \mu\right), 
\end{split}
\ee
where $T_{\text{RS}}(M, \mathcal{E})$ is the Ray-Singer torsion, in the case of Chern-Simons and 
\be
F_{BF}[A,B]= -T_{\text{RS}}(M, \mathcal{E})- \rho(M)\ln \mu. 
\ee
in the case of $BF$.

 Since 
the Weyl anomaly arises because the operations of regularizing and taking the trace
do not commute, the
anomaly ${ \cal A}^{\rm W}$ in a theory which is not classically Weyl invariant may be defined as in \eqref{A}. 
Recall, although classically a free $p$-form is equivalent to its dual $\tilde{p}$-form, where $\tilde{p}=d-2-p$, this correspondence breaks down quantum mechanically \cite{Duff:1980qv}. In the case of p-forms, there is a difference between the partition function of a $p$-form and its dual and hence between $\mathcal{A}^W$ and its dual
\cite{Duff:1980qv}: In arbitrary even dimensions
\be\label{tracediff}
\int_X {\bf \cal A}^{\rm W}(A_p)-\int_X{\bf \cal A}^{\rm W}(A_{\tilde p})=(-1)^p\frac{1}{2}(p- \tilde p)\chi(X).
\ee
Such {\it quantum inequivalence} of $p$-forms and their duals also appears in the context of {\it duality anomalies} shown in  \autoref{dualityanom}.  The $\rho$-characteristic also appears in the oscillitary mode sum but is exactly cancelled by the zero modes. 
 There is no duality anomaly in odd dimensions but we noted
\be\label{azero}\begin{split}
{\cal A}^{p,\tilde{p}}_{\text{zero}, \mu} & =\ln(\mu)\times  \left\{ \begin{array}{cc} \tfrac{(-1)^{p}}{2}(\tilde p-p) \chi(X) &\text{even}\\[5pt](-1)^{p} \rho(X) &\text{odd} \end{array} \right.
 \end{split}
\ee
so, in this sense, $(-1)^{p} \rho(X)$ play the role in odd dimensions that Weyl anomaly does in even.  For conformal field theories  in even dimensions the $a$-theorem \cite{Cardy:1988cwa,Osborn:1989td,Komargodski:2011vj} is governed by the coefficient of $\chi$. So  one might be tempted to think that in odd dimensions the $F$-theorem \cite{Myers:2010xs, Casini:2011kv, Jafferis:2011zi, Fei:2015oha} would be governed by the coefficent of $\rho$.  This is not apparent in the $F$-theorem literature, which  is, however,  confined to spheres. Perhaps the free energy of non-spherical manifolds will throw more light on this.

In the case of Kaluza-Klein, the vanishing of the duality anomaly in odd dimensions means that, when compactified to an even dimension, the contribution of the massless modes must be exactly cancelled by the infinite tower of massive KK modes, as is the case for the axial anomaly  \cite{Duff:1982wm,Duff:1982gj}. This has interesting implications for string/M theory: assuming that by virtue of its odd dimensionality M-theory has no anomaly\footnote{The R-R 1-form potential derives from the metric in $D=11$, so really this argument assumes duality anomaly freedom for gravity.}, the $D=10$ Type IIA string requires the infinite tower of massive D0-branes, in other words, the eleventh dimension of M-theory!

 \section*{Acknowledgements}

We grateful to Philip Candelas and Xenia del la Ossa for helpful discussions and comments on the manuscript.  LB is supported by the Leverhulme Research Project Grant RPG-2018-329.  MJD is grateful to Marlan Scully for his hospitality in the Institute for Quantum Science and Engineering, Texas A\&M University, and to the Hagler Institute for Advanced Study at Texas A\&M for a Faculty Fellowship. The work of MJD was supported in part by the STFC under rolling grant ST/P000762/1. SN is supported by STFC grant ST/T000686/1.

\appendix

\FloatBarrier
\section{$B_4$ coefficients for $p$-forms}\label{B4}
\FloatBarrier
Here we summarise some of the Weyl anomaly coefficient data for $p$-forms, see for example \cite{Christensen:1978md} for more details. The Weyl anomaly is given by
\be
\begin{split}
(4\pi)^2{\cal A}^{\rm W}&=cF-aG+eR^2+f\nabla^2R+gR{}~{}^*R\\
\end{split}
\ee
where
\be
G=R_{\mu\nu\rho\sigma}R^{\mu\nu\rho\sigma}-4R_{\mu\nu}R^{\mu\nu}+R^2
\ee
\be
F=C^{\mu\nu\rho\sigma}C_{\mu\nu\rho\sigma}
\ee
and
where $C_{\mu\nu\rho\sigma}$ is the Weyl tensor. 

\begin{table}[h!]
$\begin{array}{lrrrrrrrrrrrr}
A_p&&&rep&\bar a&\bar c&\bar e&\bar f&\bar g\\
&&&&\\
A_0~~~~~~~~~~~~~~~~~~~~~~~~~~~~~~~~~~~~~~~~~~&&&(0,0)&2&6&10&24&0\\
A_1&&&(1/2,1/2)&128&84&20&-24&0\\
A_2{}^+&&&(1,0)&-54&78&10&-48&120\\
A_2{}^-&&&(0,1)&-54&78&10&-48&-120\\
A_3&&&(1/2,1/2)&128&84&20&-24&0\\
A_4&&&(0,0)&2&6&0&24&0\\
&&&&&\\
E=A_2-2A_1+2A_0&&&360&-360&0&0&0&0\\
T=A_2{}^+-A_2{}^-&&&0&0&0&0&0&240\\
(E\pm T)/2=A_2{}^\pm -A_1+A_0&&&180&-180&0&0&0&\pm 120\\
A_0-A_4&&&0&0&0&0&0&0\\
A_1-A_3&&&0&0&0&0&0&0\\
\end{array}$
\caption{$p$-form $A_p$ anomaly ${\rm SO}(4)$ representations in $d=4$. Here $E$ denotes the Euler characteristic combination and $T$ the Hirzebruch signature combination. In the final two rows the na\"ively dual potentials are compared to emphasise their equivalence.  Note, $\bar x = x/720$.  }
\label{4} 
\end{table}


\begin{table}[h!]
$\begin{array}{lrrrrrrrrrrrrrrrrrr}
B_p&&&rep&a&c&e&f\\
&&&&\\
B_0=A_0&&&1&2&6&10&24\\
B_1=A_1-A_0&&&3&126&78&10&-48\\
B_2=A_2-A_1+A_0&&&3&-234&78&10&-48\\
B_3=A_3-A_2+A_1-A_0&&&1&362&6&10&24\\
B_4=A_4-A_3+A_2-A_1+A_0&&&0&-360&0&0&0\\
&&&&&\\

B_3-B_0=E&&&0&-360&0&0&0\\
B_2-B_1=-E&&&0&-360&0&0&0\\
B_4=-E&&&0&-360&0&0&0\\
\end{array}$
\caption{$p$-form $A_p$ anomaly massive $\SO(3)$ representations $B_p$ in $d=4$. The final three rows display the $(p,   \tilde p)= (2,1)$, $(p,   \tilde p)= (3, 0)$ and $(p,  \tilde p)= (4, -1)$ massive Weyl anomaly inequivalences, respectively. (Here, $p<0$ represents a formal form with vanishing anomaly coefficients.)}
\label{3} 
\end{table}

\begin{table}[h!]
$\begin{array}{lrrrrrrrrrrrrrr}
C_p&rep&a&c&e&f&g\\
&&&&\\
C_0=A_0&1&2&6&10&24&0\\
C_1=A_1-2A_0&2&124&72&0&-72&0\\
C_2=A_2-2A_1+3A_0&1&-358&6&10&24&0\\
C_3=A_3-2A_2+3A_1-4A_0&0&720&0&0&0&0\\
C_4=A_4-2A_3+3A_2-4A_1+5A_0&0&-1080&0&0&0&0&\\
&&&&&\\
C_2-C_0=E&0&-360&0&0&0&0\\
C_3=-2E&0&-720&0&0&0&0\\
C_4=3E&0&-1080&0&0&0&0&\\
(C_2-C_0)^{\pm}=\tfrac12(E\pm T)&0&-180&0&0&0&\pm 120\\
\end{array}$
\caption{$p$-form $A_p$  anomaly massless  $\SO(2)$ representations $C_p$ in $d=4$. The final four  rows display the $(p, \tilde p)= (2, 0)$, $(p, \tilde p)= (3, -1)$,  $(p, \tilde p)= (4, -2)$ massless Weyl anomaly inequivalences and the  splitting into (anti)self-dual parts.}
\label{2} 
\end{table}

Note massless  representations on $Y=X \times S^1$ with Betti numbers $b_k$ correspond to massive representations on $X $ with Betti numbers $c_k$
\be(b_0,b_1,b_2,b_3,b_4,b_5)=(c_0,c_1+c_0, c_2+c_1, c_3+c_2, c_4+c_3,c_4)\ee
\begin{equation}\label{bet1}
\begin{array}{lllllllllll}
&p&\\
&0&b_0&=\phantom{-}c_0\\
&1&b_1-2b_0&=-c_0+c_1\\
&2& b_2-2b_1+3b_0&=\phantom{-}c_0-c_1+c_2\\
&3& b_3-2b_2+3b_1-4b_0&=-c_0+c_1-c_2+c_3\\
&4&b_4-2b_3+3b_2-4b_1+5b_0&=\phantom{-}c_0-c_1+c_2-c_3+c_4\\
\end{array}
\end{equation}
where we see  the  $4$-forms  have  $\rho(Y)$ and $\chi(X)$ degrees of freedom, respectively. 

Since $\rho(Y)=\chi(X)$, the massive Weyl anomaly inequivalences in $d=4$
\[
\int_{X} {\bf \cal A}^{\rm W}(A_2)-\int_X{\bf \cal A}^{\rm W}(A_1)=\chi(X),
\]
\[
\int_{X} {\bf \cal A}^{\rm W}(A_3)-\int_X{\bf \cal A}^{\rm W}(A_0)=-\chi(X),
\]
\[
\int_{X} {\bf \cal A}^{\rm W}(A_4)=\chi(X),
\]
where in general
\[\int_{X} {\bf \cal A}^{\rm W}(A_p)
=\sum_{k=0}^{p}(-1)^k c_{p-k},
\]
follow from the zero-modes\footnote{Since the zero-mode  and  oscillatory contributions cancel in odd dimensions, one could equivalently use the oscillatory modes.} contributions to the vanishing massless inequivalences in $d=5$, when compactified on a circle 
\[
\int_{Y} {\bf \cal A}^{\rm W}(A_2)\Big|_{\text{zero}}-\int_Y{\bf \cal A}^{\rm W}(A_1)\Big|_{\text{zero}}=\rho(Y),
\]
\[
\int_{Y} {\bf \cal A}^{\rm W}(A_3)\Big|_{\text{zero}}-\int_Y{\bf \cal A}^{\rm W}(A_0)\Big|_{\text{zero}}=-\rho(Y),\]
\[
\int_{Y} {\bf \cal A}^{\rm W}(A_5)\Big|_{\text{zero}}=\rho(Y),
\]
where in general
\[\int_{Y} {\bf \cal A}^{\rm W}(A_p)\Big|_{\text{zero}}=\sum_{k=0}^{p}(-1)^k(k+1) b_{p-k}.\]


\providecommand{\href}[2]{#2}\begingroup\raggedright\endgroup

\end{document}